\documentclass{aastex63}

\usepackage{comment}
\usepackage{multirow}
\usepackage{booktabs}
\definecolor{darkviolet}{rgb}{0.58, 0.0, 0.83}

\submitjournal{ApJ}

\shorttitle{Dark Pixels}
\shortauthors{Jin et al.}
\graphicspath{{./}{figures/}}

\begin{document}

\title{(Nearly) Model-Independent Constraints on the Neutral Hydrogen Fraction in the Intergalactic Medium at $z\sim 5 - 7$ Using Dark Pixel Fractions in Ly$\alpha$ and Ly$\beta$ Forests}

\author[0000-0002-5768-738X]{Xiangyu Jin}
\affiliation{Steward Observatory, University of Arizona, 933 N. Cherry Ave., Tucson, AZ 85719, USA}

\author[0000-0001-5287-4242]{Jinyi Yang}
\thanks{Strittmatter Fellow}
\affiliation{Steward Observatory, University of Arizona, 933 N. Cherry Ave., Tucson, AZ 85719, USA}

\author[0000-0003-3310-0131] {Xiaohui Fan}
\affiliation{Steward Observatory, University of Arizona, 933 N. Cherry Ave., Tucson, AZ 85719, USA}

\author[0000-0002-7633-431X]{Feige Wang}
\thanks{NASA Hubble Fellow}
\affiliation{Steward Observatory, University of Arizona, 933 N. Cherry Ave., Tucson, AZ 85719, USA}

\author[0000-0002-2931-7824]{Eduardo Ba\~nados}
\affiliation{{Max Planck Institut f\"ur Astronomie, K\"onigstuhl 17, D-69117 Heidelberg, Germany}}

\author{Fuyan Bian}
\affiliation{European Southern Observatory, Alonso de Córdova 3107, Casilla 19001, Vitacura, Santiago 19, Chile}

\author[0000-0003-0821-3644]{Frederick B. Davies}
\affiliation{Max-Planck-Institut f\"{u}r Astronomie, K\"{o}nigstuhl 17, D-69117 Heidelberg, Germany}

\author[0000-0003-2895-6218]{Anna-Christina Eilers}\thanks{NASA Hubble Fellow}
\affiliation{MIT Kavli Institute for Astrophysics and Space Research, 77 Massachusetts Ave., Cambridge, MA 02139, USA}

\author[0000-0002-6822-2254]{Emanuele Paolo Farina}
\affiliation{Gemini Observatory, NSF’s NOIRLab, 670 N A’ohoku Place, Hilo, Hawai'i 96720, USA}

\author{Joseph F. Hennawi}
\affiliation{Department of Physics, Broida Hall, University of California, Santa Barbara, CA 93106-9530, USA}
\affiliation{Leiden Observatory, Leiden University, P.O. Box 9513, NL-2300 RA Leiden, The Netherlands}

\author[0000-0001-9879-7780]{Fabio Pacucci}
\affil{Center for Astrophysics $\vert$ Harvard \& Smithsonian,
Cambridge, MA 02138, USA}
\affil{Black Hole Initiative, Harvard University,
Cambridge, MA 02138, USA}

\author{Bram Venemans}
\affiliation{Leiden Observatory, Leiden University, P.O. Box 9513, NL-2300 RA Leiden, The Netherlands}

\author{Fabian Walter}
\affiliation{Max-Planck-Institut f\"{u}r Astronomie, K\"{o}nigstuhl 17, D-69117 Heidelberg, Germany}







\correspondingauthor{Xiangyu Jin}
\email{xiangyujin@arizona.edu}



\begin{abstract}

Cosmic reionization was the last major phase transition of hydrogen from neutral to highly ionized in the intergalactic medium (IGM). Current observations show that the IGM is significantly neutral at $z>7$, and largely ionized by $z\sim5.5$. However, most methods to measure the IGM neutral fraction are highly model-dependent, and are limited to when the volume-averaged neutral fraction of the IGM is either relatively low ($\overline{x}_{\rm HI} \lesssim 10^{-3}$) or close to unity ($\overline{x}_{\rm HI}\sim 1$). In particular, the neutral fraction evolution of the IGM at the critical redshift range of $z=6-7$ is poorly constrained. We present new constraints on $\overline{x}_{\rm HI}$ at $z\sim5.1-6.8$, by analyzing deep optical spectra of $53$ quasars at $5.73<z<7.09$.
We derive model-independent upper limits on the neutral hydrogen fraction based on the fraction of “dark” pixels identified in the Lyman $\alpha$ (Ly$\alpha$) and Lyman $\beta$ (Ly$\beta$) forests, without any assumptions on the IGM model or the intrinsic shape of the quasar continuum. They are the first model-independent constraints on the IGM neutral hydrogen fraction at $z\sim6.2-6.8$ using quasar absorption measurements. Our results give upper limits of $\overline{x}_{\rm HI}(z=6.3) < 0.79\pm0.04$ (1$\sigma$), $\overline{x}_{\rm HI} (z=6.5) < 0.87\pm0.03$ (1$\sigma$), and $\overline{x}_{\rm HI} (z=6.7) < 0.94^{+0.06}_{-0.09}$ (1$\sigma$). The dark pixel fractions at $z>6.1$ are consistent with the redshift evolution of the neutral fraction of the IGM derived from the Planck 2018. 

\end{abstract}



\section{Introduction}

Cosmic reionization was the epoch that started when UV photons from the first luminous sources ionized neutral hydrogen in the intergalactic medium (IGM) and ended the dark ages. Reionization was the last major phase transition of hydrogen in the IGM, influencing almost every baryon in the Universe. Determining when and how the reionization happened can help us to understand early structure formation and the properties of the first luminous sources in the Universe. The optical depth measured from the cosmic microwave background provides an integrated constraint on reionization, and the \textit{Planck} 2018 results infer a mid-point redshift of reionization is $z_{\rm re}=7.7\pm0.8$ \citep{Planck2020AA}. However, the detailed temporal evolution of the IGM neutral fraction, as well as its spatial variation, during the reionization era require other measurements from discrete astrophysical sources. 




The redshift evolution of the IGM neutral fraction during the reionization can be constrained by various observations. The Ly$\alpha$ and Ly$\beta$ effective optical depth measurements suggest that the IGM is highly ionized (volume-averaged IGM neutral fraction $\overline{x}_{\rm HI}\lesssim10^{-4}$) at $z\sim5.5$, while the tail end of reionization likely extends to as low as $z\sim 5.3$ \cite[e.\,g.,\,][]{Fan2006AJ,Becker2015MNRAS,Bosman2018MNRAS,Eilers2018ApJ,Eilers2019ApJ,Yang2020ApJb,Bosman2021MNRAS}.
At $z\gtrsim 6$, the emergence of complete Gunn-Peterson troughs in quasar spectra indicates a rapid increase in the neutral fraction of the IGM. At the same time, the quasar Ly$\alpha$ and Ly$\beta$ forests become saturated and their optical depth is no longer sensitive to the ionization state of the IGM. 
Close to the mid-point of reionization, the Gunn-Peterson optical depth is high enough to have strong off-resonance scattering in the form of IGM damping wings in the quasar proximity zone \citep{ME1998ApJ}. Damping wing measurements indicates the IGM is significantly neutral at $z\sim 7.1 - 7.6$ \cite[$\overline{x}_{\rm HI}\sim0.2-0.7$, ][]{Greig2017MNRAS,Banados2018Natur,Davies2018ApJ,Greig2019MNRAS,Wang2020ApJ,Yang2020ApJa,Greig2022MNRAS}. 
This leaves a gap in the IGM neutral fraction measurements between $z\sim 6  - 7$, a critical period in the reionization history when the IGM is likely experiencing the most rapid evolution. 

Apart from Ly$\alpha$ effective optical depth and IGM damping wings, high-redshift quasars can provide another constraints on $\overline{x}_{\rm HI}$: (1) The covering fraction of ``dark" pixels, present in the Ly$\alpha$ and Ly$\beta$ forests, can constrain $\overline{x}_{\rm HI}$ as model-independent upper limits \citep{Mesinger2010MNRAS,McGreer2011MNRAS,McGreer2015MNRAS}. \citet{McGreer2015MNRAS} show that $\overline{x}_{\rm HI}<0.04+0.05$ at $z=5.6$ (1$\sigma$), and $\overline{x}_{\rm HI}<0.06+0.05$ at $z=5.8$ (1$\sigma$); (2) The length distribution of long ``dark" gaps in Ly$\alpha$ and Ly$\beta$ forests can provide model-dependent constraints on $\overline{x}_{\rm HI}$ by comparing with predictions from reionization models \citep{Mesinger2010MNRAS}. \citet{Zhu2021ApJ} suggest that the dark gap statistics in Ly$\alpha$ forests favors late reionization models in which reionization ends below $z\sim6$, and \citet{Zhu2022ApJ} constrain $\overline{x}_{\rm HI}<0.05,0.17,$ and $0.29$ at $z=5.55,5.75,$ and $5.95$ from the length distribution of dark gaps in Ly$\alpha$ and Ly$\beta$ forests; (3) Mean free path of ionizing photons measured from composite quasar spectra can also be used to constrain $\overline{x}_{\rm HI}$ by comparing mean free paths with predicted results of reionization models \citep{Worseck2014MNRAS,Becker2021MNRAS}. Mean free paths measured in \citet{Becker2021MNRAS} favor late reionization models in which $\overline{x}_{\rm HI}=0.2$ at $z=6$; And (4) the size of quasar proximity zones can infer $\overline{x}_{\rm HI}$ \cite[e.\,g.,\,][]{Fan2006AJ,Carilli2010ApJ,Calverley2011MNRAS,Venemans2015ApJ,Eilers2017ApJ}, though the results are dependent on quasar lifetimes.

The process of reionization can also be constrained by high-$z$ galaxies observations through various methods: 
(1) The fraction of Ly$\alpha$ emitters (LAEs) in the broad-band selected Lyman break galaxies
\cite[e.\,g.,\,][]{Stark2010MNRAS,Pentericci2011ApJ,Schenker2014ApJ}; 
(2) The clustering (angular correlation function) of LAEs \cite[e.\,g.,\,][]{SM2015MNRAS,Ouchi2018PASJ}; 
(3) The distribution of Ly$\alpha$ equivalent width of LAEs \cite[e.\,g.,\,][]{Mason2018ApJ,Hoag2019ApJ,Mason2019MNRAS,Jung2020ApJ};
And (4) the evolution of Ly$\alpha$ luminosity functions \cite[e.\,g.,\,][]{Konno2014ApJ,Konno2018PASJ,Itoh2018ApJ,Morales2021ApJ}.

Almost all the methods of measuring the neutral fraction of the IGM discussed above are {\em model-dependent}: they rely on a number of assumptions including models of IGM density distributions, reconstruction of quasar intrinsic spectra, quasar lifetime, or intrinsic evolution of Ly$\alpha$ emission in galaxies. 
In contrast, the dark pixel method gives the least model-dependent constraints on $\overline{x}_{\rm HI}$. This method was first proposed in \citet{Mesinger2010MNRAS}, which uses the covering fraction of dark pixels of $\sim$3~Mpc size as simple upper limits on $\overline{x}_{\rm HI}$, since both pre-overlap and post-overlap neutral patches in the IGM can cause dark pixels. This method thus hardly relies on the modeling of the intrinsic emission of the quasar nor on IGM models. The dark pixel method only assumes the size of neutral patches is bigger than 3~Mpc, therefore it can be used as a nearly model-independent probe of reionization. 
The drawback is that without assuming a specific IGM density distribution, the dark pixel fraction is strictly an upper limit on the neutral fraction. 
Using the covering fraction of dark pixels, \citet{McGreer2015MNRAS} have derived stringent constraints on $\overline{x}_{\rm HI}$ at $z<6$ based a sample of 22 quasars at $5.73<z<6.42$. 

In this work, we expand these studies by using a much larger sample of $53$ quasars, and expand the redshift range to $5.73<z<7.09$.
This allows us to derive new constraints on $\overline{x}_{\rm HI}$ at $5.1<z<6.8$ by measuring the covering fraction of dark pixels. 
In particular, it provides reliable upper limits of of $\overline{x}_{\rm HI}$ at $z>6.2$ for the first time. This paper is organized as follows: we present the dataset used in our analysis in Section \ref{sec:data}, the dark pixel method in Section \ref{sec:method}, results and discussion in Section \ref{sec:results}, and our conclusion in Section \ref{sec:conclusion}. Throughout this paper, we adopt a flat $\Lambda$CDM cosmology with cosmological parameters $H_0=70~{\rm km\;\!s^{-1}\;\!Mpc^{-1}}$ and $\Omega_{\rm M}=0.3$.

\section{Data Preparation}\label{sec:data}
\begin{figure*}
    \centering
    \includegraphics[width=\textwidth]{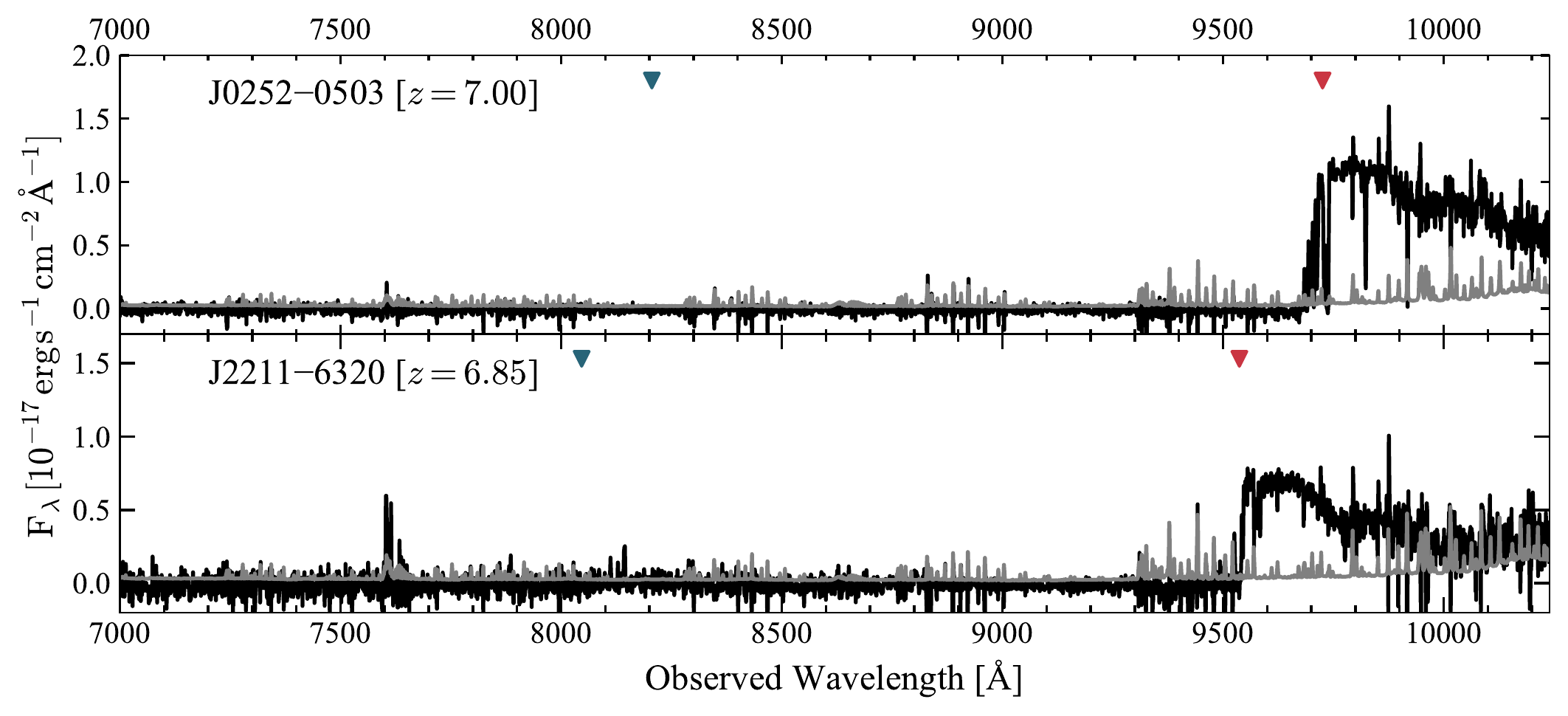}
    \caption{VLT/X-Shooter spectra of J0252--0503 and J2211--6320 in the observed wavelength. The original spectrum is shown by the black line, and spectral uncertainty is in gray. The observed wavelengths of Ly$\alpha$ and Ly$\beta$ emission lines are denoted by the red and blue triangles, respectively. Both spectra are smoothed with a median filter of $5$ pixels for better visualization.}
    \label{fig:0252_2211_xshooter}
\end{figure*}

The spectra of the $53$ quasars used in this work include most of the spectra presented in \citet{McGreer2011MNRAS,McGreer2015MNRAS} and in \citet{Yang2020ApJb}. 
The quasar sample in \citet{McGreer2011MNRAS,McGreer2015MNRAS} includes 29 spectra of 22 quasars at $5.73<z<6.42$, obtained with  Keck II Telescope/Echellette Spectrograph and Imager (ESI), Magellan Baade Telescope/Magellan Echellette (MagE) Spectrograph, Multi-Mirror Telescope (MMT)/Red Channel Spectrograph, and Vergy Large Telescope (VLT)/X-Shooter. 
The quasar sample in \citet{Yang2020ApJb} includes 35 spectra of 32 quasars at $6.31<z<7.00$ obtained with VLT/X-Shooter, Keck II/DEep Imaging Multi-Object Spectrograph (DEIMOS), Keck I/Low Resolution Imaging Spectrometer (LRIS), Gemini/Gemini Multi-Object Spectrographs (GMOS), Large Binocular Telescope (LBT)/Multi-Object Double CCD Spectrographs (MODS), and MMT/BINOSPEC. For the data reduction of these spectra, we refer the reader to \citet{McGreer2011MNRAS,McGreer2015MNRAS} and \citet{Yang2020ApJb} for more details. In addition to the spectra in \citet{McGreer2011MNRAS,McGreer2015MNRAS} and \citet{Yang2020ApJb}, we have also included new VLT/X-Shooter spectra for quasars J0252--0503 ($z=7.00$) and J2211--6320 ($z=6.84$) in our study, both taken in 2019, and an archival VLT/X-Shooter spectrum for a quasar J1120+0641 at $z=7.085$ \citep{Mortlock2011Natur}, taken in 2011 \citep{Barnett2017AA}. For the new VLT/X-Shooter spectra of J0252-0503 and J2211-6320, we perform the data reduction for bias subtracting, flat-fielding, and flux calibration with \texttt{PypeIt} \citep{Prochaska2020JOSS,Prochaska2020zndo}, following the standard thread\footnote{\url{https://pypeit.readthedocs.io/en/release/step-by-step.html}}. We present these two VLT/X-Shooter spectra in Figure \ref{fig:0252_2211_xshooter}. 

\begin{figure}
    \centering
    \includegraphics[width=0.5\textwidth]{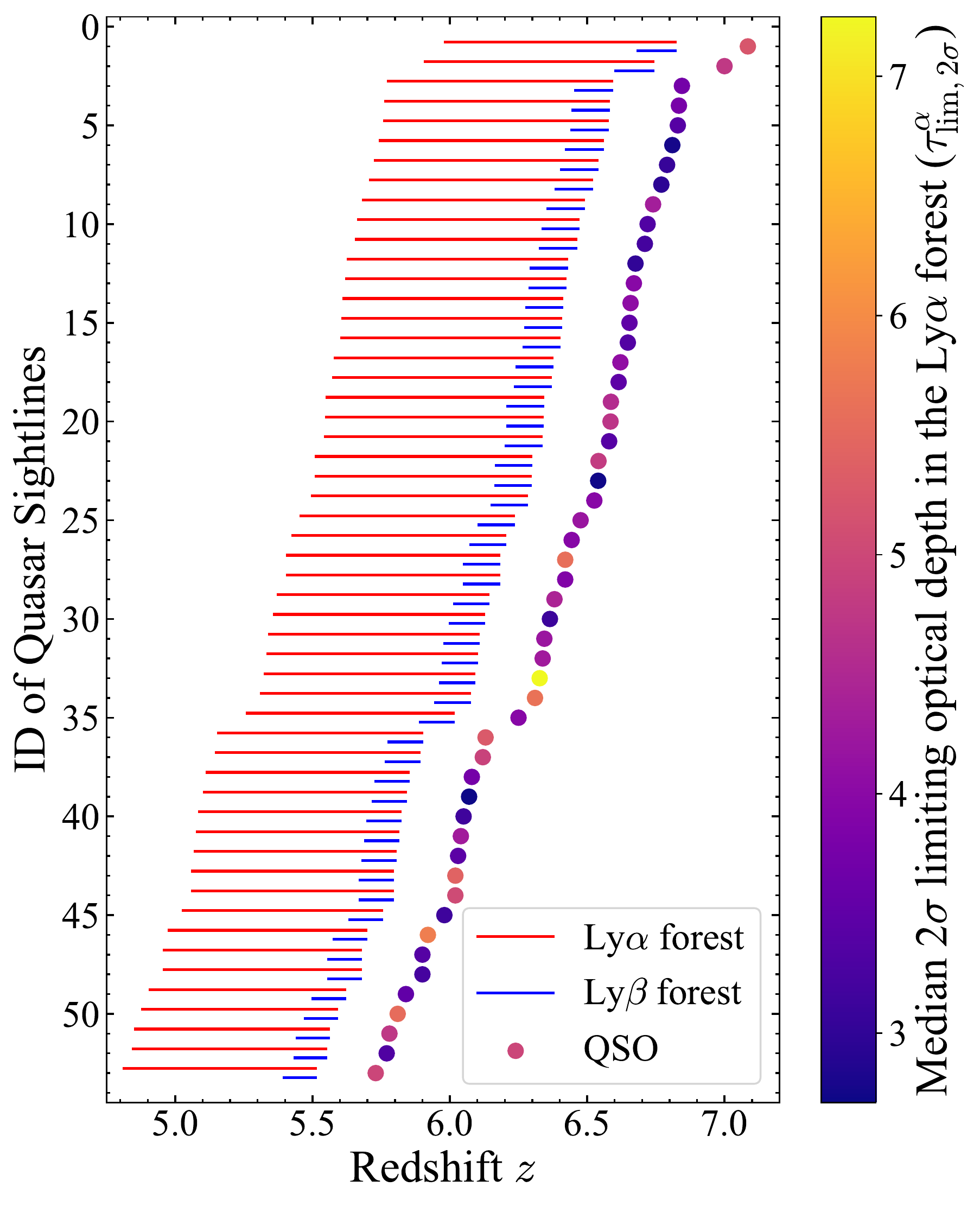}
    \caption{Redshift distribution of all quasars (circles) used in this study, and the redshift ranges of their Ly$\alpha$ (red lines) and Ly$\beta$ forests (blue lines) used in our analysis. The corresponding optical spectroscopic information of a quasar sightline ID can be found in Table \ref{tab:Data_Information}. The median $2\sigma$ limiting optical depth in the Ly$\alpha$ forest ($\tau^{\alpha}_{\rm lim,2\sigma}$) of each quasar sightline is color-coded, calculated in Section \ref{sec:method}, showing the average depth in the Ly$\alpha$ forest. A higher median $\tau^{\alpha}_{\rm lim,2\sigma}$ denotes that this quasar sightline is able to probe more opaque patches in the IGM.}
    \label{fig:qso_z_dist}
\end{figure}

We summarize the optical spectroscopy of all $53$ quasars in Table \ref{tab:Data_Information}, in descending order of redshift. We show the redshift distribution of all quasars, and the redshift range of Ly$\alpha$ and Ly$\beta$ forests used in our data pixel fraction analysis in Figure \ref{fig:qso_z_dist}.

For objects with multiple spectra, we use the histogram method to stack these spectra to improve the signal-to-noise ratio: we first set a common wavelength grid, based on the spectrum with lowest spectral resolution among all the spectra of the same object. Then we use the inverse variance weighting to calculate the flux, and the spectral uncertainty of each pixel on the common wavelength grid to obtain a stack spectrum. 

For the range of the Ly$\alpha$ forest used in our analysis, we choose the blue cutoff at $1050~{\rm \AA}$ in the rest-frame to exclude the possible emission from \ion{O}{6} $\lambda$1033 \citep{Bosman2021MNRAS}. We choose a red cutoff  at $1176~{\rm \AA}$ in the rest-frame to avoid possible contamination from the quasar proximity zone\footnote{A rest-frame wavelength of $1176~{\rm \AA}$ is corresponding to $14.9~$proper Mpc from a $z=5.73$ quasar and to $11.3~$proper Mpc from a $z=7.09$ quasar.}.
For the wavelength coverage of the Ly$\beta$ forest, we select a blue cutoff at $975~{\rm \AA}$ in the rest-frame to avoid contamination from Lyman $\gamma$ forests. We also match the red cut of the Ly$\beta$ forest to the same absorption redshift as the red cut of the Ly$\alpha$ forest (i.e., $1176~{\rm \AA}\times \lambda_{\rm Ly\beta}/\lambda_{\rm Ly \alpha}$ in the rest-frame, where $\lambda_{\rm Ly \alpha}=1215.7~{\rm \AA}$ and $\lambda_{\rm Ly\beta}=1025.7~{\rm \AA}$ are the rest wavelengths of Ly$\alpha$ and Ly$\beta$), resulting in a wavelength range of $975-992.2~{\rm \AA}$.

To minimize the contamination of strong sky emission lines (mainly ${\rm OH}$ emission) in our analysis, we first apply a median filter of 3 pixels to smooth the spectrum, and then mask pixels which are above $3\sigma$ in both flux density and spectral uncertainty than the smoothed spectrum. We reject pixels with SNR $<-5$ caused by oversubtraction of sky. 
Due to the sky ${\rm O}_2$ emission \citep{Osterbrock1996PASP}, we also mask the SNR $>2$ pixels in the observed range of $8620-8680~{\rm \AA}$ range. This may exclude real transmission spikes, but SNR $>2$ pixels within this region cannot be identified as transmission spikes precisely, based on the current data quality.
\cite{Eilers2019ApJ} showed that the metal absorption line contribution is negligible at $z\sim6$, and we thus do not correct them in our analysis. 

\begin{figure*}
    \centering
    \includegraphics[width=0.5\textwidth]{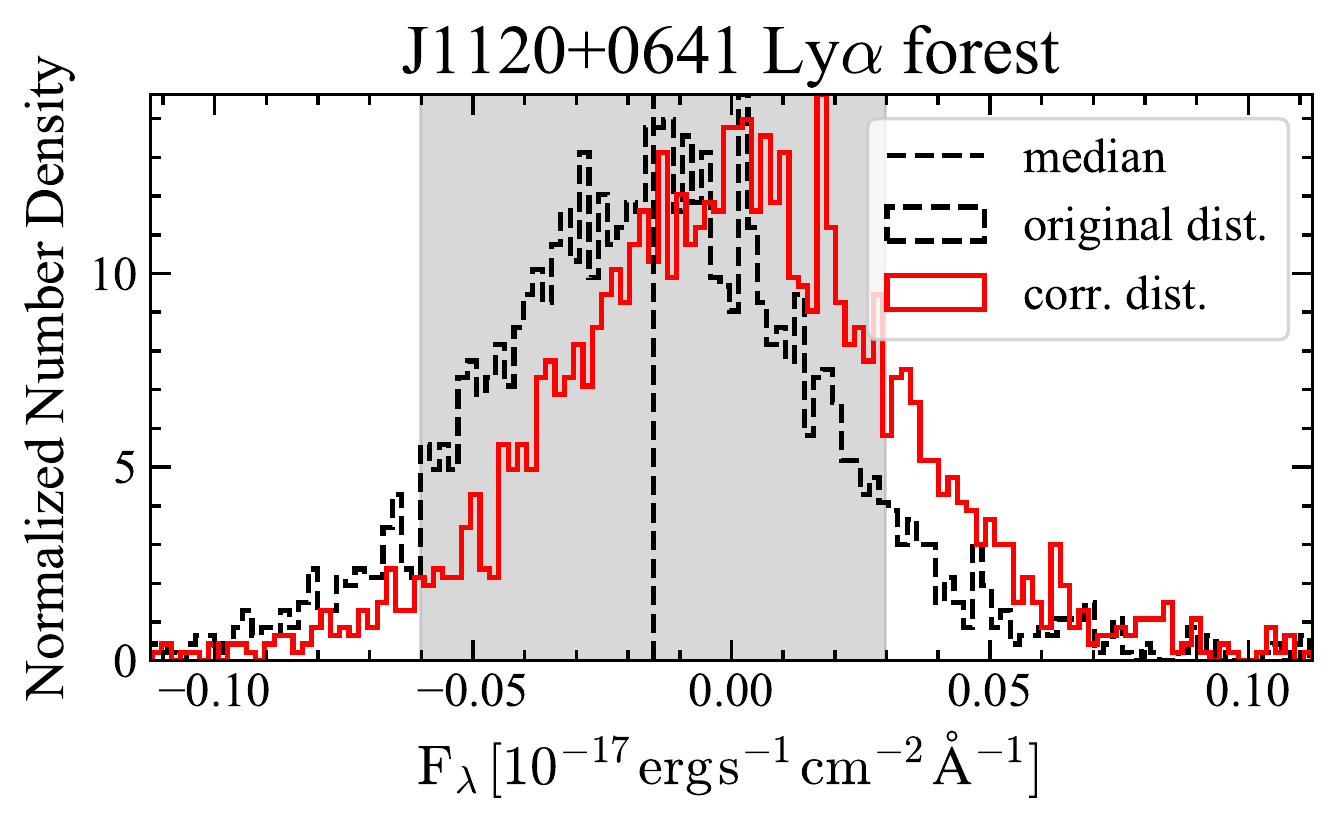}
    \caption{To correct ``zero" flux level in VLT/X-Shooter spectra, we apply the median flux derived from sigma clipped pixels to the VLT/X-Shooter spectra. The original flux distribution of pixels in the Ly$\alpha$ forest from the J1120+0641 VLT/X-Shooter spectrum is shown by the black histogram. The median derived from the sigma clipping is denoted by the vertical black dashed line, and the 2$\sigma$ range from sigma clipping is in the grey shaded region. The flux distribution corrected by the sigma-clipped median is shown by the red histogram.}
    \label{fig:1120_flux_distribution}
\end{figure*}

For some VLT/X-Shooter spectra in our study (especially at the high-redshift end), the sky background level is not precisely subtracted, resulting in a ``zero" flux offset in these spectra. This flux floor is removed empirically as follows: 
after skyline masking, we first investigate the flux distribution of pixels in the Ly$\alpha$ forest. Then we perform $2\sigma$ sigma-clipping on the pixel flux until convergence of the mean and the median flux is achieved. Figure \ref{fig:1120_flux_distribution} shows the flux distribution of pixels in the Ly$\alpha$ forest as the black histogram from the J1120+0641 VLT/X-Shooter spectrum. 
The median flux from the sigma-clipped pixels is denoted by the vertical dashed line, and the $2\sigma$ range from the sigma clipping is represented by the grey shaded region. We use the median flux of sigma-clipped pixels to correct the zero flux level for all VLT/X-Shooter spectra of quasars. The average flux correction in transmitted flux is $\sim0.08\%-4\%$.

\startlongtable
\begin{deluxetable*}{ccccc}\centering 
\tabletypesize{\scriptsize}
\tablecaption{Information of Quasar Optical Spectroscopy}
\tablewidth{2\columnwidth}
\tablehead{\colhead{ID} & 
\colhead{Name} & \colhead{$z$} & \colhead{Telescope/Instrument} & \colhead{Median $\tau^{\alpha}_{\rm lim,2\sigma}$}}
\colnumbers
\startdata
1 & J1120+0641  &  7.09 & VLT/X-Shooter & 5.23 \\
2 & J0252--0503  &  7.00 & VLT/X-Shooter & 4.76\\
3 & J2211--6320  &  6.84 & VLT/X-Shooter & 3.73  \\
4 & J0020--3653  &  6.83 & VLT/X-Shooter & 3.77    \\
5 & J0319--1008  &  6.83 & Gemini/GMOS & 3.40   \\
6 & J0411--0907  &  6.81 & LBT/MODS & 2.78   \\
7 & J0109--3047  &  6.79 & VLT/X-Shooter & 3.05  \\
8 & J0218+0007  &  6.77 & Keck/LRIS & 2.95  \\
9 & J1104+2134  &  6.74 & Keck/LRIS & 4.33  \\
10 & J0910+1656  &  6.72 & Keck/LRIS & 3.34   \\
11 & J0837+4929  &  6.71 & LBT/MODS & 3.19  \\
 &            &       & MMT/BINOSPEC &  \\
12 & J1048--0109  &  6.68 & VLT/X-Shooter & 3.00  \\
13 & J2002--3013  &  6.67 &  Gemini/GMOS & 3.98  \\
14 & J2232+2930  &  6.66 & VLT/X-Shooter & 4.04  \\
15 & J1216+4519  &  6.65 & Gemini/GMOS  & 3.49 \\
 &            &        & Keck/LRIS &    \\
 &           &        & LBT/MODS &   \\
16 & J2102--1458  &  6.65 & Keck/DEIMOS & 3.36  \\
17 & J0024+3913  &  6.62 & Keck/DEIMOS & 4.08  \\
18 & J0305--3150  &  6.61 & VLT/X-Shooter & 3.48 \\
19 & J1526--2050  &  6.59 & Keck/DEIMOS & 4.56  \\
20 & J2132+1217  &  6.59 & Keck/DEIMOS & 4.71 \\
21 & J1135+5011  &  6.58 & MMT/BINOSPEC & 3.39  \\
22 &J0226+0302  &  6.54 & Keck/DEIMOS & 4.81  \\
23 & J0148--2826  &  6.54 & Gemini/GMOS & 2.73  \\
24 & J0224--4711  &  6.53 & VLT/X-Shooter & 3.98  \\
25 & J1629+2407  &  6.48 & Keck/DEIMOS & 4.20  \\
26 & J2318--3113  &  6.44 & VLT/X-Shooter & 3.93  \\
27 & J1148+5251  &  6.42 & Keck/ESI  & 5.60  \\
28 & J0045+0901  &  6.42 & Keck/DEIMOS & 3.89 \\
29 & J1036--0232  &  6.38 & Keck/DEIMOS & 4.45  \\
30 & J1152+0055  &  6.36 & VLT/X-Shooter & 3.11  \\
31 & J1148+0702  &  6.34 & VLT/X-Shooter & 4.20  \\
32 & J0142--3327  &  6.34 & VLT/X-Shooter & 4.28  \\
33 & J0100+2802  &  6.33 & VLT/X-Shooter & 7.25  \\
34 & J1030+0524  &  6.31 & Keck/ESI & 5.64   \\
 &            &       & VLT/X-Shooter &    \\
35 & J1623+3112  &  6.25 & Keck/ESI  & 3.95  \\
36 & J1319+0950  &  6.13 & VLT/X-Shooter & 5.27 \\
37 & J1509--1749  &  6.12 & Magellan/MagE & 4.92 \\
 &            &       & VLT/X-Shooter &   \\
38 & J0842+1218  &  6.08 & Keck/ESI  & 3.75  \\
39 & J1630+4012  &  6.07 & MMT/Red Channel Spectrograph & 2.71 \\
40 & J0353+0104  &  6.05 & Keck/ESI  & 3.15  \\
41 & J2054--0005  &  6.04 & Magellan/MagE& 4.29  \\
42 & J1137+3549  &  6.03 & Keck/ESI  & 3.45 \\
43 & J0818+1722  &  6.02 & MMT/Red Channel Spectrograph   & 5.42  \\
  &           &       & VLT/X-Shooter \\
44 & J1306+0356  &  6.02 & Keck/ESI & 5.04  \\
 &           &       & VLT/X-Shooter &    \\
45 & J0841+2905  &  5.98 & Keck/ESI  & 3.13 \\
46 & J0148+0600  &  5.92 & VLT/X-Shooter & 5.80  \\
47 & J1411+1217  &  5.90 & Keck/ESI  & 3.37  \\
48 & J1335+3533  &  5.90 & Keck/ESI  & 3.22  \\
49 & J0840+5624  &  5.84 & Keck/ESI  & 3.52 \\
50 & J0836+0054  &  5.81 & Keck/ESI     & 5.56   \\
 &            &       & MMT/Red Channel Spectrograph      &   \\
 &            &       & VLT/X-Shooter &   \\
51 & J1044--0125  &  5.78 & Magellan/MagE& 4.74  \\
52 & J0927+2001  &  5.77 & Keck/ESI & 3.32  \\
53 & J1420--1602  &  5.73 & Magellan/MagE& 4.96 \\
\enddata
\tablecomments{(1) ID of quasar sightlines, in descending order of redshift. (2) Name of quasar. (3) Redshift. (4) Instrument used to obtain the spectrum. (5) The median of $2\sigma$ limiting optical depth in the Ly$\alpha$ forest, on a pixel scale of $3.3~{\rm cMpc}$. If there are multiple spectra of one object, the listed $2\sigma$ limiting optical depth is given for the stacked spectrum.}
\end{deluxetable*}\label{tab:Data_Information}

\section{Methods}\label{sec:method}

\begin{figure}
    \centering
    \includegraphics[width=\columnwidth]{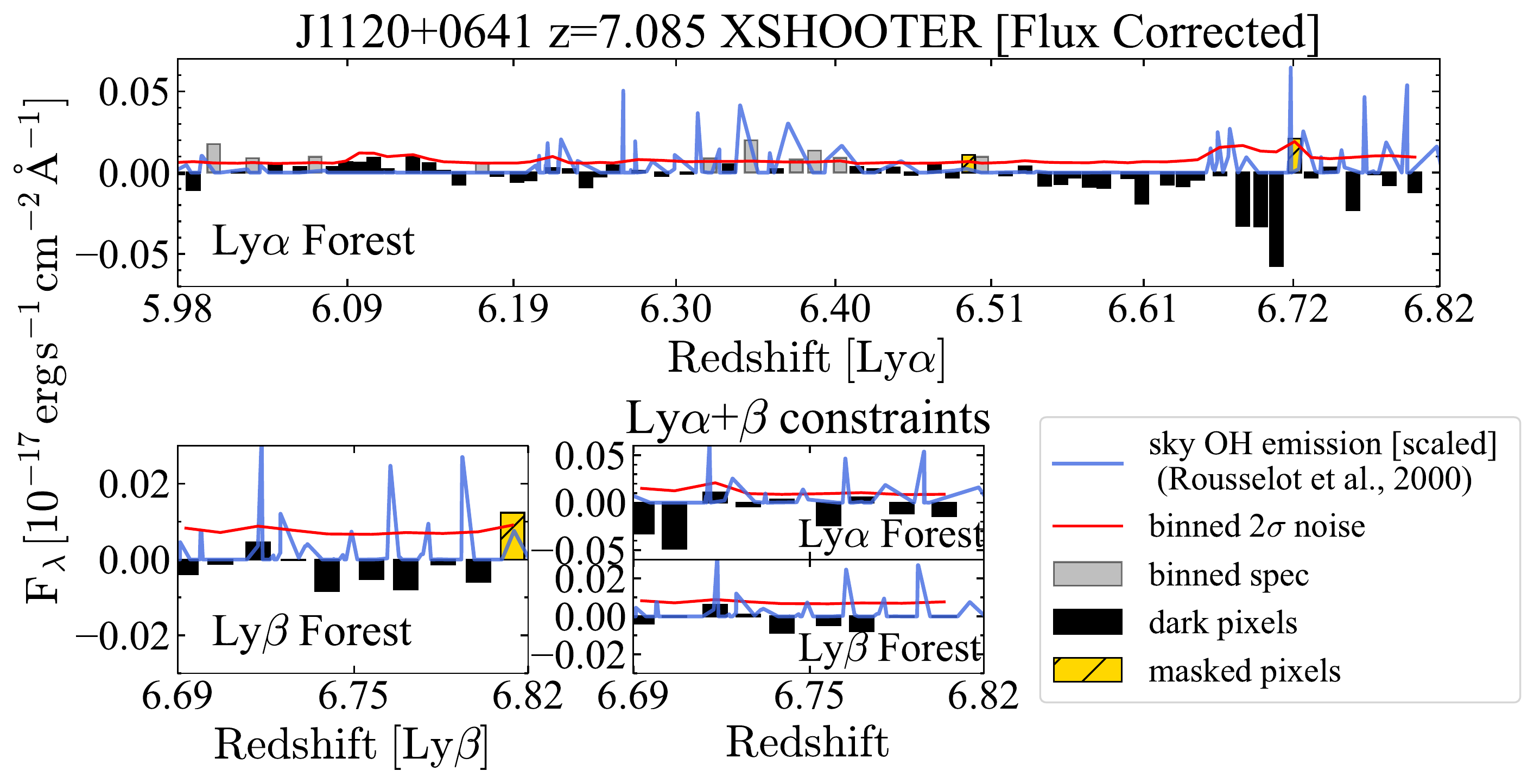}
    \caption{J1120+0641 binned spectrum in the Ly$\alpha$ forests (upper panel), Ly$\beta$ forests (lower left panel), and the redshift overlapping regions in Ly$\alpha$ and Ly$\beta$ forests (lower middle panel). The size of each binned pixel is $3.3~$cMpc. The binned $2\sigma$ spectral uncertainty is shown as the red line. The binned flux density is presented in bars, and identified ``dark" pixels are shown in black. The sky OH emission is shown by the blue line \citep{Rousselot2000AA}, and our spectral binning method effectively puts the majority of pixels between sky OH emission lines. Yellow hatched pixels are manually masked during our analysis, as those fluxes are plausibly from residual sky OH emission. The zero flux level is corrected by the $2\sigma$ clipping median of all pixels in the Ly$\alpha$ forest (i.\,e.,\,the vertical dashed line shown in Figure \ref{fig:1120_flux_distribution}).
    }
    \label{fig:binned_spec_correction}
\end{figure}

To improve the dynamic range of the spectrum, we follow a similar method as the method described in \citet{McGreer2011MNRAS} to perform spectral binning. The size of each binned pixel is $3.3~{\rm Mpc}$ in the comoving distance (i.\,e.,\,$3.3~{\rm cMpc}$), following \citet{McGreer2011MNRAS,McGreer2015MNRAS}. To avoid any contamination caused by residual sky lines in the spectrum, we first identify the local minima in the $1\sigma$ spectral uncertainty in the Ly$\alpha$ and Ly$\beta$ forests with \texttt{argrelextrema} in \texttt{Scipy} \citep{2020SciPy-NMeth} and an order of 3, which identifies those local minima that are less than their 3 neighboring pixels in the spectral uncertainty. We place the $3.3~{\rm cMpc}$ pixels centered at those local minima, until the interval between any two adjacent pixels is less than $3.3~{\rm cMpc}$. 
We then use the inverse variance weighting to calculate the flux and the spectral uncertainty of each $3.3~{\rm cMpc}$ binned pixel. As an example, Figure \ref{fig:binned_spec_correction} shows the J1120+0641 binned spectrum, corrected with the $2\sigma$ clipping median of all pixels.  
Before calculating the covering fraction of dark pixels, we perform a visual inspection on every binned spectrum by comparing it with near-infrared sky OH emission lines \citep{Rousselot2000AA}. We manually mask any bright pixel plausibly caused by sky emission at $z>6.3$ in the binned spectrum. These manually masked ``sky" pixels are denoted by yellow hatched pixels in Figure \ref{fig:binned_spec_correction}. 

We adopt a flux threshold method to identify ``dark pixels" in the binned spectra, following \citet{McGreer2011MNRAS}. Pixels with flux density less than $2\sigma$, where $\sigma$ is the binned spectral uncertainty, are identified as ``dark pixels". These dark pixels are denoted by black bars in Figure \ref{fig:binned_spec_correction}. \citet{McGreer2011MNRAS,McGreer2015MNRAS} introduced an alternative definition of the ``dark" pixel fraction, as twice the fraction of pixels with negative flux. Since the ``dark" pixels intrinsically have zero flux, there is a probability of $0.5$ for them to scatter below $0$ flux. We do not adopt this negative flux pixel definition, because this method requires an extremely precise background subtraction, which is difficult to achieve for the highest redshift quasar spectra in this study due to the sky background (see Section \ref{sec:data}). 
As dark pixel fractions are used as upper limits on $\overline{x}_{\rm HI}$, we then calculate the ratio of total number of dark pixels to the total number of pixels of all quasar lines of sight as the dark pixel fraction
, within a redshift bin of $\Delta z=0.2$, for both the Ly$\alpha$ transition (from $z=5.2$ to $z=6.8$) and the Ly$\beta$ transition (from $z=5.4$ to $z=6.8$). In each redshift bin, we use jackknife statistics to derive the $1\sigma$ uncertainty in the dark pixel fraction. 

Apart from the individual constraints from Ly$\alpha$ and Ly$\beta$ forests, we also derive a combined dark pixel fraction in Ly$\alpha$ and Ly$\beta$ forests from their redshift overlapping regions \citep{McGreer2011MNRAS}. 
For this combined dark pixel fraction, we stack the spectral uncertainty in Ly$\alpha$ and Ly$\beta$ forests at the same redshift using the inverse variance weighting, and utilize the stacked spectral uncertainty to put 3.3~cMpc pixels at local minima. The corresponding binned spectrum is shown on the lower middle panel in Figure \ref{fig:binned_spec_correction}. In this constraint, a pixel is ``dark" only if its flux density is below $2\sigma$ binned spectral uncertainty in the both Ly$\alpha$ and Ly$\beta$ transitions. The redshift range used to calculate the dark pixel fraction for this combining constraints from Ly$\alpha$ and Ly$\beta$ forests is the same as the redshift range used to calculate the dark pixel fraction in Ly$\beta$ forests. 

We perform a continuum fitting of the original spectrum by assuming a broken power-law with a break at the rest-frame 1000$~{\rm \AA}$ \citep{Shull2012ApJ}. We use the least square method to fit the spectrum within $1245-1285~{\rm \AA}$ and $1310-1380~{\rm \AA}$\footnote{For J0024+3913 and J2132+1217, only the wavelength range of $1245-1285~{\rm \AA}$ is used in the continuum fitting, since the spectrum in 1310-1380~${\rm \AA}$ is noisy.} in the rest-frame with a fixed spectral index ($\alpha_{\lambda}$) of $-1.5$, following \citet{Yang2020ApJb}, and derive the normalization of the power-law continuum. We then calculate the continuum flux at rest-frame $\lambda>1000~{\rm \AA}$ with the best-fit normalization and a spectral index of $-1.5$. At rest-frame $\lambda<1000~{\rm \AA}$, we switch the spectral index to $\alpha_{\lambda}=-0.59$ to calculate the continuum flux. 

We calculate the $2\sigma$ limiting optical depth  $\tau_{\rm lim,2\sigma}=-{\rm ln}(2\sigma/F_{\rm cont})$ for each $3.3~$cMpc~pixel
where $2\sigma$ is the binned uncertainty on a pixel size of $3.3~$cMpc, and $F_{\rm cont}$ is the best-fit continuum flux. A higher limiting optical depth indicates that the pixel can place stronger constraints on the neutral hydrogen fraction in the IGM. We present the median limiting optical depth in the Ly$\alpha$ forest of the binned spectra (on a pixel scale of $3.3~{\rm cMpc}$) in Table \ref{tab:Data_Information}. For pixels in Ly$\beta$ forests, we correct their limiting optical depth by subtracting the effective optical depth of foreground Ly$\alpha$ forests, using the measured Ly$\alpha$ effective optical depth relations in \citet[][for foreground Ly$\alpha$ forests at $z<5.3$]{Fan2006AJ} and in \citet[][for foreground Ly$\alpha$ forests at $5.3<z<6.0$]{Yang2020ApJb}. When calculating the dark pixel fraction, we exclude $\tau_{\rm lim,2\sigma}<2.5$ for Ly$\alpha$ pixels (i.\,e.,\,$\tau^{\alpha}_{\rm lim,2\sigma}<2.5$), 
as those pixels do not have enough sensitivity to probe the neutral hydrogen. Considering the Ly$\alpha$ and Ly$\beta$ transitions have different oscillator strengths, the corresponding cut in a limiting optical depth for Ly$\beta$ pixels will be $\tau^{\beta}_{\rm lim,2\sigma}<2.50/2.19\sim1.14$, assuming a conversion factor of $2.19$ between Ly$\alpha$ and Ly$\beta$ effective optical depth \citep{Fan2006AJ}. Furthermore, we re-calculate the dark pixel fraction only with $\tau^{\alpha}_{\rm lim,2\sigma}>4.5$ pixels (corresponding to $\tau^{\beta}_{\rm lim,2\sigma}>2.05$ for Ly$\beta$ pixels) to constrain the neutral hydrogen fraction with high quality pixels.

\section{Results and Discussion}\label{sec:results}
\begin{figure}
    \centering
    \includegraphics[width=0.49\columnwidth]{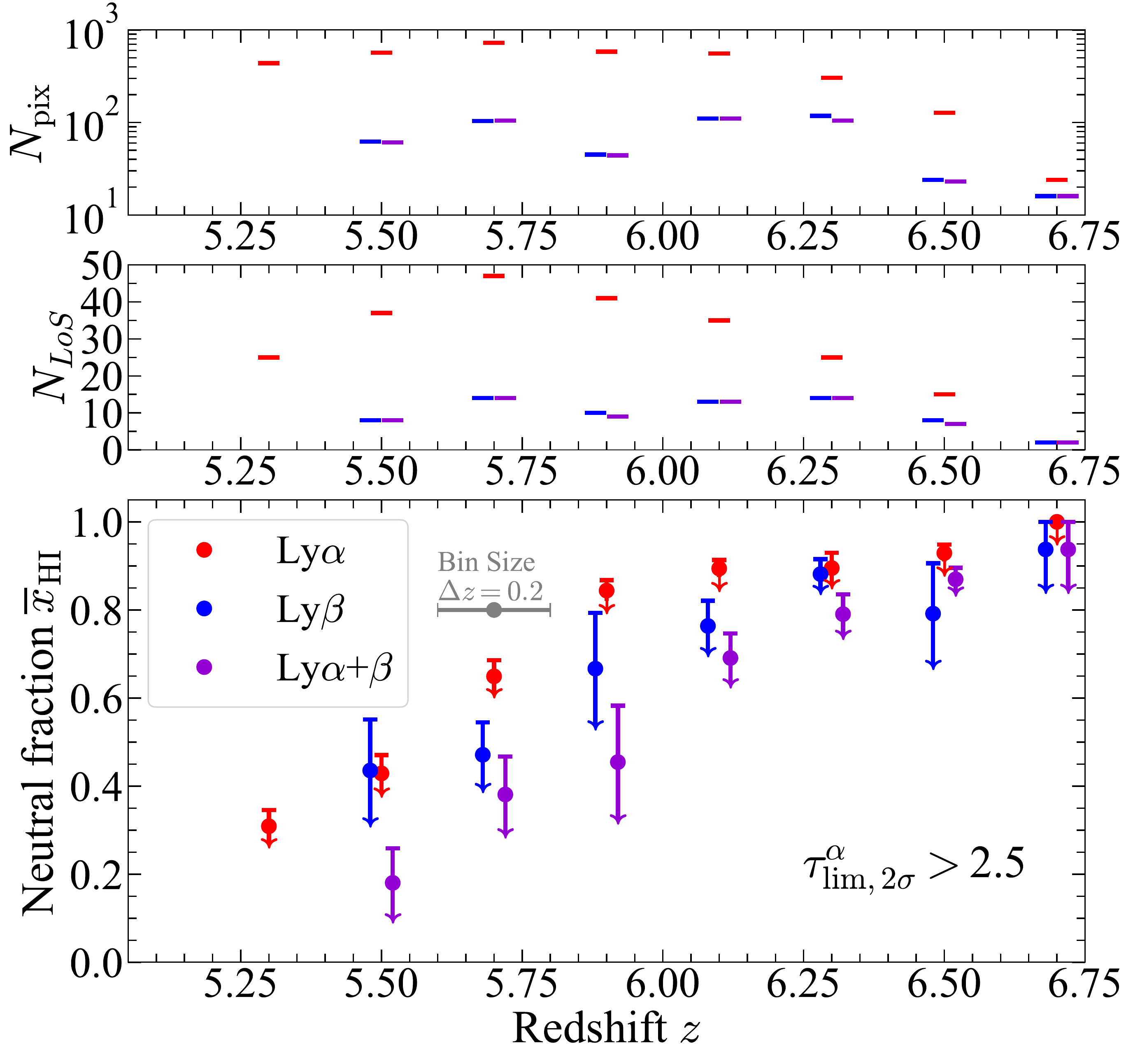}
    \includegraphics[width=0.49\columnwidth]{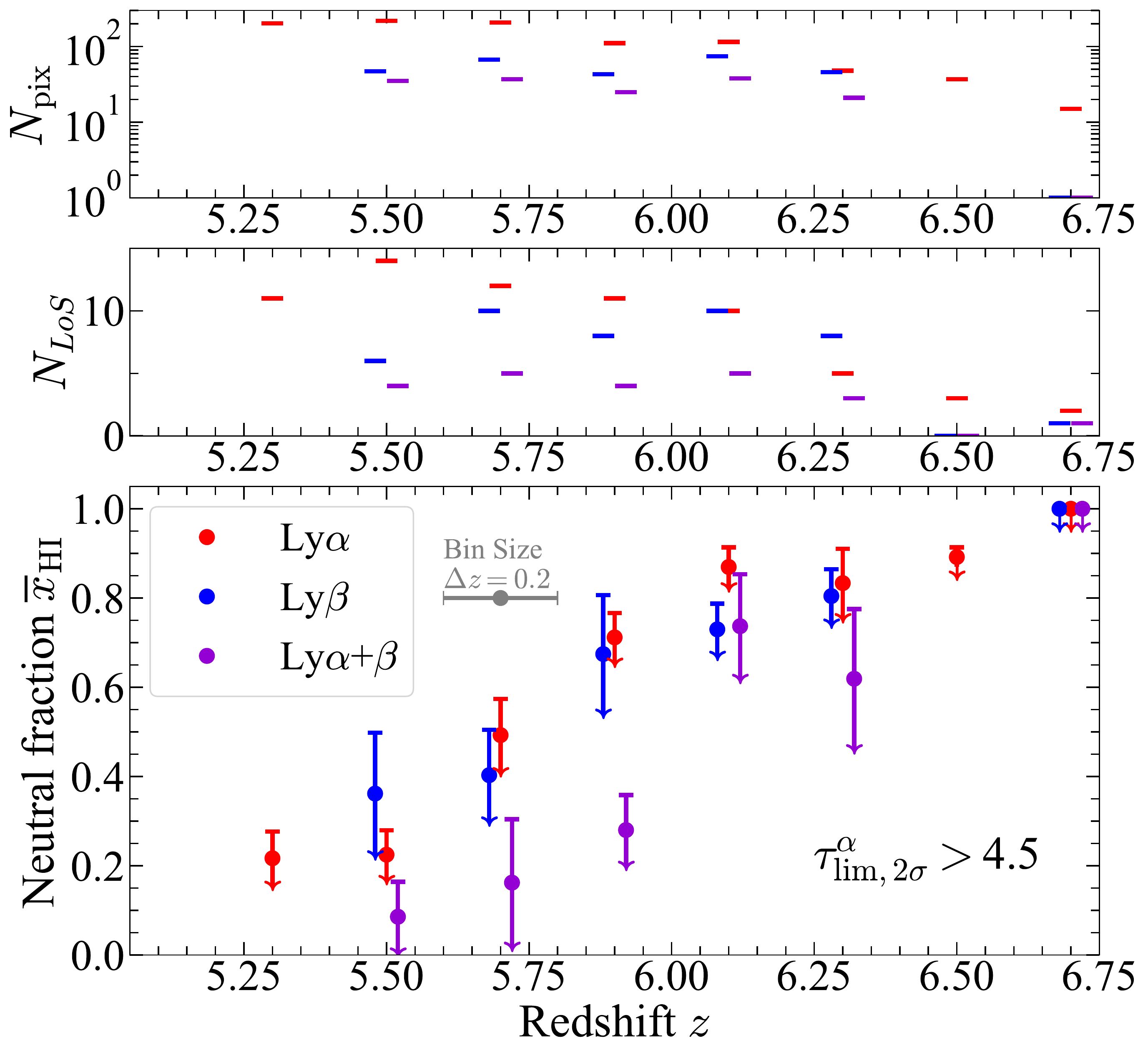}
    \caption{\textit{Upper Panels --} Number of $3.3~$cMpc pixels used to calculate the dark pixel fraction in the Ly$\alpha$ (red), Ly$\beta$ (blue) forests and the combined dark fraction in the Ly$\alpha$ and Ly$\beta$ forests (purple) in each bin of $\Delta z=0.2$ in redshift; \textit{Middle Panels --} Number of quasar lines of sight available in each bin of $\Delta z=0.2$ in redshift; \textit{Bottom Panels --} Upper limits on the volume-averaged IGM neutral fraction $\overline{x}_{\rm HI}$ derived from the dark pixel fractions in the Ly$\alpha$ (red) and Ly$\beta$ (blue) forests as a function of redshift. The combined dark pixel fractions from the Ly$\alpha$ and Ly$\beta$ forests are denoted by purple markers. We adopt the same redshift bin ranges when calculating the dark pixel fractions derived from Ly$\alpha$ forests (red), Ly$\beta$ forests (blue), and combined Ly$\alpha$ and Ly$\beta$ forests (purple), but here we apply a small offset ${\Delta}z=0.02$ in the figure for better visualization. The three panels on the left show the results with $3.3~$cMpc pixels of which the limiting optical depth is greater than $2.5$; Three panels on the right are the results with $3.3~$cMpc pixels of which the limiting optical depth is greater than $4.5$. Note there is no $\tau^{\alpha}_{\rm lim,2\sigma}>4.5$ pixel in the Ly$\beta$ forest in the $z=6.5$ bin. 
    }
    \label{fig:nh_corr}
\end{figure}

We present the redshift evolution of upper limits on $\overline{x}_{\rm HI}$ from dark pixels in Figure \ref{fig:nh_corr}.
The number of $\tau^{\alpha}_{\rm lim,2\sigma}>2.5$ pixels in each $\Delta z=0.2$ bin in redshift, the number of lines of sight in each redshift bin, and the upper limits derived from $\tau^{\alpha}_{\rm lim,2\sigma}>2.5$ pixels are shown in the left panel, and the results of $\tau^{\alpha}_{\rm lim,2\sigma}>4.5$ are shown in right panel. In both two panels, the combined dark pixel fractions from Ly$\alpha$ and Ly$\beta$ forests give the most stringent upper limits on $\overline{x}_{\rm HI}$. From the combined dark pixel fraction derived from $\tau^{\alpha}_{\rm lim,2\sigma}>2.5$ pixels in Ly$\alpha$ and Ly$\beta$ forests, the upper limit on the neutral hydrogen is $18\%\pm8\%$ at $z=5.5$, and it increases to $69\%\pm 6\%$ at $z=6.1$, $79\%\pm4\%$ at $z=6.3$, $87\%\pm3\%$ at $z=6.5$, and $94\%_{-9\%}^{+6\%}$ at $z=6.7$. 
By adopting a higher limiting optical depth cut at $4.5$ than at $2.5$, the number of available pixels and the number of available quasar lines of sight drop significantly in each $\Delta z=0.2$ redshift bin. Furthermore, the upper limit on $\overline{x}_{\rm HI}$ becomes tighter at $z<6$. The upper limit on $\overline{x}_{\rm HI}$ is $9\%\pm8\%$ at $z=5.5$, $16\%\pm14\%$ at $z=5.7$, and $28\%\pm8\%$ at $z=5.9$. At $z>6$, dark pixel fractions derived from $\tau^{\alpha}_{\rm lim,2\sigma}>4.5$ pixels increase significantly, which can be caused by the rapid evolution in the IGM Ly$\alpha$ and Ly$\beta$ effective optical depth \cite[e.g.,\,][]{Yang2020ApJb}. However, this rapid increase in the combined Ly$\alpha$ and Ly$\beta$ dark pixel fraction can also be associated with a small data sample, as only $5$ quasar lines of sight are available with $\tau^{\alpha}_{\rm lim,2\sigma}>4.5$ pixels at $z=6.1$. Furthermore, at $z>6$, dark pixel fractions derived from $\tau^{\alpha}_{\rm lim,2\sigma}>4.5$ pixels do not necessarily provide more stringent upper limits on $\overline{x}_{\rm HI}$ than those derived from $\tau^{\alpha}_{\rm lim,2\sigma}>2.5$ pixels.
At $z>6.3$, the number of available $\tau^{\alpha}_{\rm lim,2\sigma}>4.5$ pixels is very limited. For example, at $z\sim6.4-6.6$ (the central redshift of the bin is $z=6.5$), there is no $\tau^{\alpha}_{\rm lim,2\sigma}>4.5$ pixel in Ly$\beta$ forests, due to the lack of high signal-to-noise ratio quasar spectra and the narrow wavelength range of Ly$\beta$ forests used in our analysis. 
We tabulate the redshift distributions of the number of quasar lines of sight, the number dark pixels, the number of pixels, and the value of dark pixel fractions in Table \ref{tab:Dark_pixel_fraction}.

\begin{deluxetable*}{cccccc|cccc}\centering 
\tabletypesize{\scriptsize}
\tablecaption{Redshift distribution of the numbers of quasar lines of sight, of dark pixels, of all pixels, and resulting dark pixel fractions}
\tablewidth{2\columnwidth}
\tablehead{
 \colhead{\multirow{4}{*}{Constraints}} & \colhead{\multirow{4}{*}{Redshift $z$}} & \multicolumn{4}{c}{$\tau^{\alpha}_{\rm lim,2\sigma}>2.5$ pixels} & \multicolumn{4}{c}{$\tau^{\alpha}_{\rm lim,2\sigma}>4.5$ pixels}  \\ \cmidrule(lr){3-6} \cmidrule(lr){7-10}
& & \colhead{$N_{\rm LoS}$} & \colhead{$N_{\rm dark}$} & \colhead{$N_{\rm pix}$} & \colhead{$f_{\rm dark}$}  & \colhead{$N_{\rm LoS}$} & \colhead{$N_{\rm dark}$} & \colhead{$N_{\rm pix}$} & \colhead{$f_{\rm dark}$} \\
 \colhead{(1)} & \colhead{(2)} & \colhead{(3)} &\colhead{(4)} &\colhead{(5)} & \colhead{(6)} & \colhead{(7)} & \colhead{(8)} & \colhead{(9)} & \colhead{(10)}}
\startdata
\multirow{8}{*}{Ly$\alpha$}& 5.3 & 25 & 135 & 437 & $ 0.31\pm0.04 $ & 11 & 44 & 203 & $ 0.22\pm0.06 $\\
& 5.5 & 37 & 244 & 569 & $ 0.43\pm0.04 $ & 14 & 49 & 218 & $ 0.22\pm0.05 $\\
& 5.7 & 47 & 472 & 727 & $ 0.65\pm0.04 $ & 12 & 102 & 207 & $ 0.49\pm0.08 $\\
& 5.9 & 41 & 492 & 583 & $ 0.84\pm0.02 $ & 11 & 79 & 111 & $ 0.71\pm0.05 $\\
& 6.1 & 35 & 498 & 557 & $ 0.89\pm0.02 $ & 10 & 100 & 115 & $ 0.87\pm0.04 $\\
& 6.3 & 25 & 273 & 305 & $ 0.90\pm0.03 $ & 5 & 40 & 48 & $ 0.83\pm0.08 $\\
& 6.5 & 15 & 118 & 127 & $ 0.93\pm0.02 $ & 3 & 33 & 37 & $ 0.89\pm0.02 $\\
& 6.7 & 2 & 24 & 24 & $ 1.00 $ & 2 & 15 & 15 & $ 1.00$\\ \hline
\multirow{7}{*}{Ly$\beta$} & 5.5 & 8 & 27 & 62 & $ 0.44\pm0.12 $ & 6 & 17 & 47 & $ 0.36\pm0.14 $\\
& 5.7 & 14 & 49 & 104 & $ 0.47\pm0.07 $ & 10 & 27 & 67 & $ 0.40\pm0.10 $\\
& 5.9 & 10 & 30 & 45 & $ 0.67\pm0.13 $ & 8 & 29 & 43 & $ 0.67\pm0.13 $\\
& 6.1 & 13 & 84 & 110 & $ 0.76\pm0.06 $ & 10 & 54 & 74 & $ 0.73\pm0.06 $\\
& 6.3 & 14 & 104 & 118 & $ 0.88\pm0.03 $ & 8 & 37 & 46 & $ 0.80\pm0.06 $\\
& 6.5 & 8 & 19 & 24 & $ 0.79\pm0.11 $ & 0 & 0 & 0 & $-$\\
& 6.7 & 2 & 15 & 16 & $ 0.94^{+0.06}_{-0.09} $ & 1 & 1 & 1 & $ 1.00$\\ \hline
\multirow{7}{*}{Combined Ly$\alpha+$Ly$\beta$} & 5.5 & 8 & 11 & 61 & $ 0.18\pm0.08 $ & 4 & 3 & 35 & $ 0.09\pm0.08 $\\
& 5.7 & 14 & 40 & 105 & $ 0.38\pm0.09 $ & 5 & 6 & 37 & $ 0.16\pm0.14 $\\
& 5.9 & 9 & 20 & 44 & $ 0.45\pm0.13 $ & 4 & 7 & 25 & $ 0.28\pm0.08 $\\
& 6.1 & 13 & 76 & 110 & $ 0.69\pm0.06 $ & 5 & 28 & 38 & $ 0.74\pm0.12 $\\
& 6.3 & 14 & 83 & 105 & $ 0.79\pm0.04 $ & 3 & 13 & 21 & $ 0.62\pm0.16 $\\
& 6.5 & 7 & 20 & 23 & $ 0.87\pm0.03 $ & 0 & 0 & 0 & $-$\\
& 6.7 & 2 & 15 & 16 & $ 0.94^{+0.06}_{-0.09} $ & 1 & 1 & 1 & $ 1.00$\\
\enddata
\tablecomments{(1) Type of dark pixel fractions. (2) Central redshift of each $\Delta z=0.2$ bin. (3) The number of quasar lines of sight that have $\tau^{\alpha}_{\rm lim,2\sigma}>2.5$ pixels in this redshift bin. (4) The number of $\tau^{\alpha}_{\rm lim,2\sigma}>2.5$ dark pixels. (5) The total number of $\tau^{\alpha}_{\rm lim,2\sigma}>2.5$ pixels. (6) Dark pixel fraction derived from $\tau^{\alpha}_{\rm lim,2\sigma}>2.5$ pixels. (7) The number of quasar lines of sight that have $\tau^{\alpha}_{\rm lim,2\sigma}>4.5$ pixels in this redshift bin. (8) The number of $\tau^{\alpha}_{\rm lim,2\sigma}>4.5$ dark pixels. (9) The total number of $\tau^{\alpha}_{\rm lim,2\sigma}>4.5$ pixels. (10) Dark pixel fraction derived from $\tau^{\alpha}_{\rm lim,2\sigma}>4.5$ pixels. 
All the errors show $1\sigma$ confidence intervals.}
\end{deluxetable*}\label{tab:Dark_pixel_fraction}

\begin{figure}
    \centering
    \includegraphics[width=0.9\columnwidth]{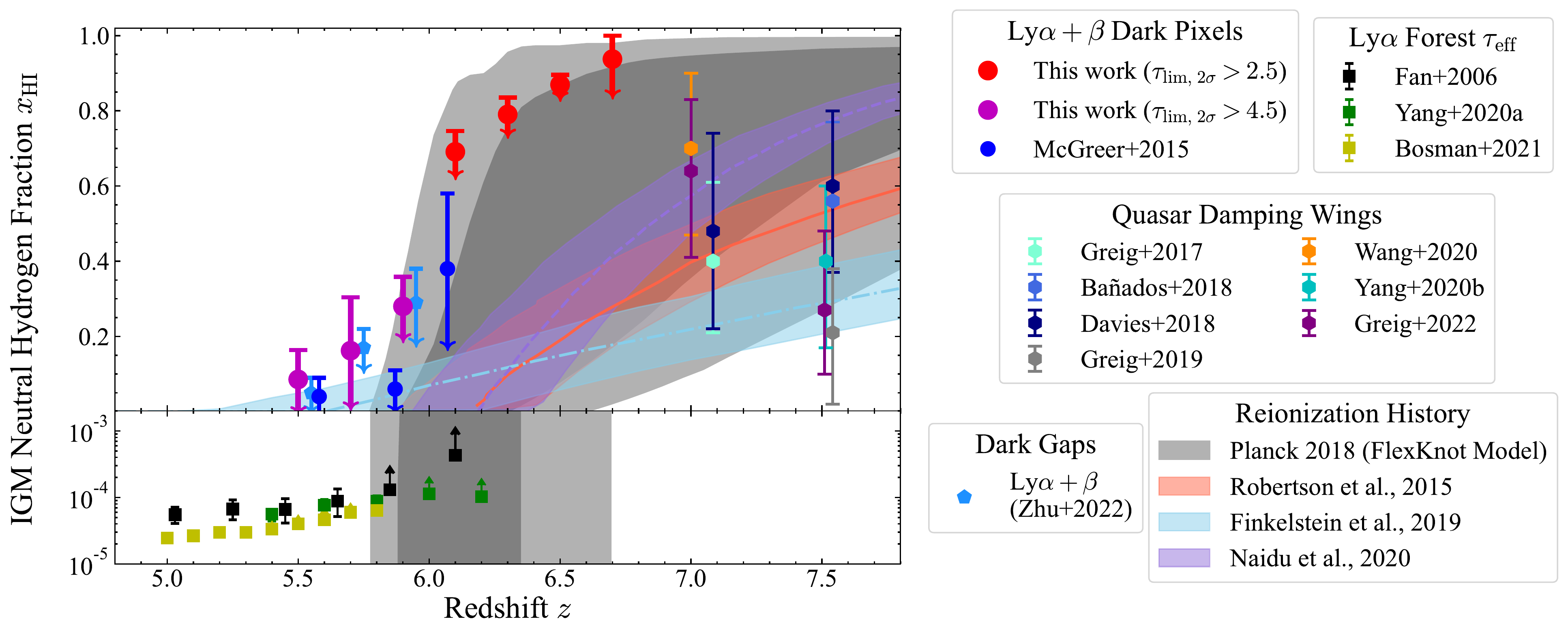}
    \caption{Constraints on the IGM neutral hydrogen fraction $\overline{x}_{\rm HI}$ from high-$z$ quasars studies and Planck 2018 results. The upper limits on $\overline{x}_{\rm HI}$ from dark pixels are in red (this work, derived from flux-corrected $\tau^{\alpha}_{\rm lim,2\sigma}>2.5$ pixels at $z>6$), magenta (this work, derived from flux-corrected $\tau^{\alpha}_{\rm lim,2\sigma}>4.5$ pixels at $z<6$), and blue \citep{McGreer2015MNRAS}. Constraints on $\overline{x}_{\rm HI}$ derived from Ly$\alpha$ effective optical depth are shown by black squares \citep{Fan2006AJ}, green squares \citep{Yang2020ApJb}, and yellow squares \citep{Bosman2021MNRAS}. The upper limits on $\overline{x}_{\rm HI}$, inferred from long dark gap length distributions in Ly$\alpha$ and Ly$\beta$ forests, are shown in blue pentagons \citep{Zhu2022ApJ}. At $z>7$, individual measurements on the neutral hydrogen fraction from quasar damping wings are denoted by hexagons \citep{Greig2017MNRAS,Banados2018Natur,Davies2018ApJ,Greig2019MNRAS,Wang2020ApJ,Yang2020ApJa}. The $1\sigma$ and $2\sigma$ reionization history derived from Planck 2018 results by assuming the FlexKnot model are denoted by the dark grey shaded region and the light grey shaded region \citep{Planck2020AA}. The colored regions display $1\sigma$ reionization histories in \citealt[][]{Robertson2015ApJ} (red), \citealt{Finkelstein2019ApJ} (blue), and \citealt{Naidu2020ApJ} (purple).}
    \label{fig:nhi_other_constraints}
\end{figure}

We show our upper limits on the IGM neutral hydrogen fraction, along with other constraints on neutral hydrogen fractions from high-redshift quasars in Figure \ref{fig:nhi_other_constraints}. Since the dark pixel fraction derived from $\tau^{\alpha}_{\rm lim,2\sigma}>4.5$ pixels can provide tighter constraints on the neutral hydrogen fraction at $z<6$, and at $z>6$ the number of $\tau^{\alpha}_{\rm lim,2\sigma}>2.5$ pixels is much higher than the number of $\tau^{\alpha}_{\rm lim,2\sigma}>4.5$ pixels, we present the dark pixel fraction derived from $\tau^{\alpha}_{\rm lim,2\sigma}>2.5$ pixels at $z>6$, denoted by red upper limits. The dark pixel fraction calculated with $\tau^{\alpha}_{\rm lim,2\sigma}>4.5$ pixels at $z<6$ are denoted by magenta upper limits. The dark pixel fraction in \citet{McGreer2015MNRAS} at $5.5<z<6.2$, derived from $22$ quasars are shown by blue upper limits. Our upper limits on the neutral fraction at $z<6$ are slightly higher than the upper limits in \citet{McGreer2015MNRAS}. The possible reasons for this difference include: (1) \citet{McGreer2015MNRAS} double the covering fraction of negative pixels as the dark pixel fraction, while the dark pixel in this work is defined by $2\sigma$ flux threshold \citep{McGreer2011MNRAS}, and (2) to avoid possible contamination from quasar proximity zones and intrinsic spectra, we adopt narrower wavelength ranges for both Ly$\alpha$ forests 
than the wavelength ranges 
used in \citet{McGreer2015MNRAS}.
We repeat the results in \citet{McGreer2015MNRAS} and test the above two factors in the resulted dark pixel fractions. We notice that the dark pixel definition (either dark pixels are defined by flux threshold or negative pixels) accounts for the major difference between our results and \citet{McGreer2015MNRAS}. Adopting a dark pixel definition of $2\sigma$ flux threshold, the combined Ly$\alpha$ and Ly$\beta$ dark pixel fractions in \citet{McGreer2015MNRAS} will become $0.16\pm0.08$ at $z=5.58$, $0.31\pm0.10$ at $z=5.87$, and $0.63\pm0.24$ at $z=6.07$, derived from $\tau^{\alpha}_{\rm lim,1\sigma}>4.5$ (corresponding $\tau^{\alpha}_{\rm lim,2\sigma}>3.8$) pixels. Although flux threshold definition gives more conservative dark pixel fractions at $z<6$, it is the only applicable method when deriving dark pixel fractions at the high-redshift end in this study, due to the strong sky emission. 


In Figure \ref{fig:nhi_other_constraints}, we show the upper limits on $\overline{x}_{\rm HI}$ from long dark gap size distributions in Ly$\alpha$ and Ly$\beta$ forests \citep{Zhu2022ApJ}, assuming a late reionization that ends at $z\lesssim5.3$ \citep{ND2020MNRAS}. Our constraints at $z<6$ are highly consistent with these upper limits derived from dark gap statistics. We also present constraints on $\overline{x}_{\rm HI}$ measured from the Ly$\alpha$ effective optical depth \citep{Fan2006AJ,Yang2020ApJb,Bosman2021MNRAS}. These $\overline{x}_{\rm HI}$ measurements from Ly$\alpha$ effective optical depth suggest the IGM is highly ionized and $\overline{x}_{\rm HI}\lesssim10^{-4}$ at $z<5.5$. At $z>6$, the Ly$\alpha$ effective optical depth measurements show that $\overline{x}_{\rm HI}\gtrsim10^{-4}$ \citep{Yang2020ApJb}. 
The IGM damping wing feature embedded in the $z>7$ quasar spectra can be used to constrain the IGM neutral fraction, based on models of the IGM morphology and quasar intrinsic emission \cite[e.\,g.,\,][]{SMH2013MNRAS}. 
In Figure \ref{fig:nhi_other_constraints}, we show several recent measurements on neutral fraction at $z>7$ from IGM damping wings in hexagons \citep{Greig2017MNRAS,Banados2018Natur,Davies2018ApJ,Greig2019MNRAS,Wang2020ApJ,Yang2020ApJa,Greig2022MNRAS}. Their medians show $\overline{x}_{\rm HI}\sim0.2-0.7$, suggesting that the IGM is significantly neutral at $z\gtrsim7$. 

The reionization history, derived from Planck 2018 results, assuming the FlexKnot model \citep{Planck2020AA}, is shown by the dark grey shaded region ($1\sigma$ confidence level), and the light grey shaded region ($2\sigma$ confidence level). The FlexKnot model reconstructs the reionization history with arbitrary number of knots, interpolates the reionization history between knots, and utilizes the Bayesian interference to marginalize the number of knots \citep{Millea2018AA}. We also include $1\sigma$ reionization histories from \citet[][red region]{Robertson2015ApJ}, \citet[][blue region]{Finkelstein2019ApJ}, and \citet[][purple region]{Naidu2020ApJ}. The ionizing budget during reionization is dominated by faint galaxies ($M_{\rm UV}>-15$) in \citet{Finkelstein2019ApJ}, while the reionization photon budget in the models of \citet{Robertson2015ApJ} and \citet{Naidu2020ApJ} is dominated by bright galaxies. 
Our upper limits on neutral hydrogen fraction at $6.3\lesssim z\lesssim6.7$ are within the $1\sigma$ reionization history (assuming FlexKnot model) from Planck 2018. However, the upper limits on $\overline{x}_{\rm HI}$ derived from dark pixels are not very efficient in distinguishing the other three reionization histories at $z>6$ shown in Figure \ref{fig:nhi_other_constraints}. This results from the limited number of quasar sight lines at $z>6.8$, and the noisy sky background in the observed wavelength of interest, leading to a small number of pixels with high limiting optical depth at $z\gtrsim6.5$. Deeper optical spectroscopy on existing $z>6.8$ quasars and more quasar lines of sight at $z>6.8$, as well as potential observations from space, are needed for future similar studies to generate more stringent constraints on $\overline{x}_{\rm HI}$ at $z>6.5$. 

\section{Conclusion}\label{sec:conclusion}
In this paper, we present the dark pixel fractions in Ly$\alpha$ and Ly$\beta$ forests of $53$ quasars at $5.73<z<7.09$. These dark pixel fractions provide the first model-independent upper limits on the volume-averaged IGM neutral fraction at $6.2<z<6.8$: $\overline{x}_{\rm HI} (z=6.3) < 0.79\pm0.04$ (1$\sigma$), $\overline{x}_{\rm HI} (z=6.5)<0.87\pm0.03$ (1$\sigma$), and $\overline{x}_{\rm HI} (z=6.7)<0.94^{+0.06}_{-0.09}$ (1$\sigma$). The dark pixel fractions at $z<6.1$ in this work are slightly higher than the dark pixel fractions presented in \citet{McGreer2015MNRAS}, due to a different definition of dark pixels used in this work and the selection of different wavelength ranges in Lyman series forests for dark pixel fraction calculation. We find that the dark pixel fractions at $z>6.2$ are consistent with the $1\sigma$ IGM neutral fraction evolution derived from the Planck 2018 results when assuming the FlexKnot model \citep{Planck2020AA}. 

The current upper limits on $\overline{x}_{\rm HI}$, derived from dark pixels, are not stringent enough to distinguish various reionization histories \cite[e.g.,\,][]{Robertson2015ApJ,Finkelstein2019ApJ,Naidu2020ApJ}. The future improvement of similar dark pixel studies requires more quasar sightlines, deeper optical spectroscopy covering Lyman series forests, and observations from space to exclude the potential contamination from sky OH emission lines.

\begin{acknowledgements}
We thank the anonymous reviewer for their constructive comments. We thank Dr.\,Ian McGreer for providing codes to reconstruct results in \citet{McGreer2015MNRAS}, and for informative discussion. XJ, JY, and XF acknowledge supports by NSF grants AST 19-08284. FW thanks the support provided by NASA through the NASA Hubble Fellowship grant $\#$HST-HF2-51448.001-A awarded by the Space Telescope Science Institute, which is operated by the Association of Universities for Research in Astronomy, Incorporated, under NASA contract NAS5-26555. ACE acknowledges support by NASA through the NASA Hubble Fellowship grant $\#$HF2-51434 awarded by the Space Telescope Science Institute, which is operated by the Association of Universities for Research in Astronomy, Inc., for NASA, under contract NAS5-26555. FP acknowledges support from a Clay Fellowship administered by the Smithsonian Astrophysical Observatory, and from the Black Hole Initiative at Harvard University, which is funded by grants from the John Templeton Foundation and the Gordon and Betty Moore Foundation. EPF acknowledges support from the international Gemini Observatory, a program of NSF’s NOIRLab, which is managed by the Association of Universities for Research in Astronomy (AURA) under a cooperative agreement with the National Science Foundation, on behalf of the Gemini partnership of Argentina, Brazil, Canada, Chile, the Republic of Korea, and the United States of America.

This work is based in part on observations made with ESO telescopes at the La Silla Paranal Observatory under program IDs 084.A-0360(A), 084.A-0390(A), 084.A-0550(A), 085.A-0299(A), 086.A-0162(A), 087.A-0607(A), 087.A-0890(A), 088.A-0897(A), 096.A-0095(A), 096.A-0418(A), 097.B-1070(A), 098.A-0444 (A), 098.A-0527(A), 098.B-0537(A), 0100.A-0446(A), 0100.A-0625(A), 0102.A-0154(A), and 0103.A-0423(A). 
Some of the data presented herein were obtained at the W. M. Keck Observatory, which is operated as a scientific partnership among the California Institute of Technology, the University of California and the National Aeronautics and Space Administration. The Observatory was made possible by the generous financial support of the W. M. Keck Foundation.
Some of the observations reported here were obtained at the MMT Observatory, a joint facility of the University of Arizona and the Smithsonian Institution.
This paper includes data gathered with the 6.5 meter Magellan Telescopes located at Las Campanas Observatory, Chile.
The paper also includes data gathered with the LBT. The LBT is an international collaboration among institutions in the United States, Italy and Germany. LBT Corporation partners are: The University of Arizona on behalf of the Arizona university system; Istituto Nazionale di Astrofisica, Italy; LBT Beteiligungsgesellschaft, Germany, representing the Max-Planck Society, the Astrophysical Institute Potsdam, and Heidelberg University; The Ohio State University, and The Research Corporation, on behalf of The University of Notre Dame, University of Minnesota and University of Virginia. 
The paper also used data based on observations obtained at the international Gemini Observatory, a program of NSF’s NOIRLab, which is managed by the Association of Universities for Research in Astronomy (AURA) under a cooperative agreement with the National Science Foundation. on behalf of the Gemini Observatory partnership: the National Science Foundation (United States), National Research Council (Canada), Agencia Nacional de Investigaci\'{o}n y Desarrollo (Chile), Ministerio de Ciencia, Tecnolog\'{i}a e Innovaci\'{o}n (Argentina), Minist\'{e}rio da Ci\^{e}ncia, Tecnologia, Inova\c{c}\~{o}es e Comunica\c{c}\~{o}es (Brazil), and Korea Astronomy and Space Science Institute (Republic of Korea). 

The authors wish to recognize and acknowledge the very significant cultural role and reverence that the summit of Maunakea has always had within the indigenous Hawaiian community. We are most fortunate to have the opportunity to conduct observations from this mountain. 

We respectfully acknowledge the University of Arizona is on the land and territories of Indigenous peoples. Today, Arizona is home to 22 federally recognized tribes, with Tucson being home to the O'odham and the Yaqui. Committed to diversity and inclusion, the University strives to build sustainable relationships with sovereign Native Nations and Indigenous communities through education offerings, partnerships, and community service.

\end{acknowledgements}

\facilities{Gemini (GMOS), Keck I (LRIS), Keck II (DEIMOS, ESI), LBT (MODS), Magellan Baade (MagE), MMT (BINOSPEC, Red Channel Spectrograph), VLT(X-Shooter)}

\software{Astropy \citep{astropy:2013,astropy:2018,astropy:2022}, Matplotlib \citep{Matplotlib2007}, NumPy \citep{Numpy2020}, PypeIt \citep{Prochaska2020JOSS,Prochaska2020zndo}, SciPy \citep{2020SciPy-NMeth}}

\bibliography{sample63}

\begin{thebibliography}{}
\expandafter\ifx\csname natexlab\endcsname\relax\def\natexlab#1{#1}\fi
\providecommand{\url}[1]{\href{#1}{#1}}
\providecommand{\dodoi}[1]{doi:~\href{http://doi.org/#1}{\nolinkurl{#1}}}
\providecommand{\doeprint}[1]{\href{http://ascl.net/#1}{\nolinkurl{http://ascl.net/#1}}}
\providecommand{\doarXiv}[1]{\href{https://arxiv.org/abs/#1}{\nolinkurl{https://arxiv.org/abs/#1}}}

\bibitem[{{Astropy Collaboration} {et~al.}(2013){Astropy Collaboration},
  {Robitaille}, {Tollerud}, {Greenfield}, {Droettboom}, {Bray}, {Aldcroft},
  {Davis}, {Ginsburg}, {Price-Whelan}, {Kerzendorf}, {Conley}, {Crighton},
  {Barbary}, {Muna}, {Ferguson}, {Grollier}, {Parikh}, {Nair}, {Unther},
  {Deil}, {Woillez}, {Conseil}, {Kramer}, {Turner}, {Singer}, {Fox}, {Weaver},
  {Zabalza}, {Edwards}, {Azalee Bostroem}, {Burke}, {Casey}, {Crawford},
  {Dencheva}, {Ely}, {Jenness}, {Labrie}, {Lim}, {Pierfederici}, {Pontzen},
  {Ptak}, {Refsdal}, {Servillat}, \& {Streicher}}]{astropy:2013}
{Astropy Collaboration}, {Robitaille}, T.~P., {Tollerud}, E.~J., {et~al.} 2013,
  \aap, 558, A33, \dodoi{10.1051/0004-6361/201322068}

\bibitem[{{Astropy Collaboration} {et~al.}(2018){Astropy Collaboration},
  {Price-Whelan}, {Sip{\H{o}}cz}, {G{\"u}nther}, {Lim}, {Crawford}, {Conseil},
  {Shupe}, {Craig}, {Dencheva}, {Ginsburg}, {Vand erPlas}, {Bradley},
  {P{\'e}rez-Su{\'a}rez}, {de Val-Borro}, {Aldcroft}, {Cruz}, {Robitaille},
  {Tollerud}, {Ardelean}, {Babej}, {Bach}, {Bachetti}, {Bakanov}, {Bamford},
  {Barentsen}, {Barmby}, {Baumbach}, {Berry}, {Biscani}, {Boquien}, {Bostroem},
  {Bouma}, {Brammer}, {Bray}, {Breytenbach}, {Buddelmeijer}, {Burke},
  {Calderone}, {Cano Rodr{\'\i}guez}, {Cara}, {Cardoso}, {Cheedella}, {Copin},
  {Corrales}, {Crichton}, {D'Avella}, {Deil}, {Depagne}, {Dietrich}, {Donath},
  {Droettboom}, {Earl}, {Erben}, {Fabbro}, {Ferreira}, {Finethy}, {Fox},
  {Garrison}, {Gibbons}, {Goldstein}, {Gommers}, {Greco}, {Greenfield},
  {Groener}, {Grollier}, {Hagen}, {Hirst}, {Homeier}, {Horton}, {Hosseinzadeh},
  {Hu}, {Hunkeler}, {Ivezi{\'c}}, {Jain}, {Jenness}, {Kanarek}, {Kendrew},
  {Kern}, {Kerzendorf}, {Khvalko}, {King}, {Kirkby}, {Kulkarni}, {Kumar},
  {Lee}, {Lenz}, {Littlefair}, {Ma}, {Macleod}, {Mastropietro}, {McCully},
  {Montagnac}, {Morris}, {Mueller}, {Mumford}, {Muna}, {Murphy}, {Nelson},
  {Nguyen}, {Ninan}, {N{\"o}the}, {Ogaz}, {Oh}, {Parejko}, {Parley}, {Pascual},
  {Patil}, {Patil}, {Plunkett}, {Prochaska}, {Rastogi}, {Reddy Janga},
  {Sabater}, {Sakurikar}, {Seifert}, {Sherbert}, {Sherwood-Taylor}, {Shih},
  {Sick}, {Silbiger}, {Singanamalla}, {Singer}, {Sladen}, {Sooley},
  {Sornarajah}, {Streicher}, {Teuben}, {Thomas}, {Tremblay}, {Turner},
  {Terr{\'o}n}, {van Kerkwijk}, {de la Vega}, {Watkins}, {Weaver}, {Whitmore},
  {Woillez}, {Zabalza}, \& {Astropy Contributors}}]{astropy:2018}
{Astropy Collaboration}, {Price-Whelan}, A.~M., {Sip{\H{o}}cz}, B.~M., {et~al.}
  2018, \aj, 156, 123, \dodoi{10.3847/1538-3881/aabc4f}

\bibitem[{{Astropy Collaboration} {et~al.}(2022){Astropy Collaboration},
  {Price-Whelan}, {Lim}, {Earl}, {Starkman}, {Bradley}, {Shupe}, {Patil},
  {Corrales}, {Brasseur}, {N{"o}the}, {Donath}, {Tollerud}, {Morris},
  {Ginsburg}, {Vaher}, {Weaver}, {Tocknell}, {Jamieson}, {van Kerkwijk},
  {Robitaille}, {Merry}, {Bachetti}, {G{"u}nther}, {Aldcroft},
  {Alvarado-Montes}, {Archibald}, {B{'o}di}, {Bapat}, {Barentsen}, {Baz{'a}n},
  {Biswas}, {Boquien}, {Burke}, {Cara}, {Cara}, {Conroy}, {Conseil}, {Craig},
  {Cross}, {Cruz}, {D'Eugenio}, {Dencheva}, {Devillepoix}, {Dietrich},
  {Eigenbrot}, {Erben}, {Ferreira}, {Foreman-Mackey}, {Fox}, {Freij}, {Garg},
  {Geda}, {Glattly}, {Gondhalekar}, {Gordon}, {Grant}, {Greenfield}, {Groener},
  {Guest}, {Gurovich}, {Handberg}, {Hart}, {Hatfield-Dodds}, {Homeier},
  {Hosseinzadeh}, {Jenness}, {Jones}, {Joseph}, {Kalmbach}, {Karamehmetoglu},
  {Ka{l}uszy{'n}ski}, {Kelley}, {Kern}, {Kerzendorf}, {Koch}, {Kulumani},
  {Lee}, {Ly}, {Ma}, {MacBride}, {Maljaars}, {Muna}, {Murphy}, {Norman},
  {O'Steen}, {Oman}, {Pacifici}, {Pascual}, {Pascual-Granado}, {Patil},
  {Perren}, {Pickering}, {Rastogi}, {Roulston}, {Ryan}, {Rykoff}, {Sabater},
  {Sakurikar}, {Salgado}, {Sanghi}, {Saunders}, {Savchenko}, {Schwardt},
  {Seifert-Eckert}, {Shih}, {Jain}, {Shukla}, {Sick}, {Simpson},
  {Singanamalla}, {Singer}, {Singhal}, {Sinha}, {Sip{H{o}}cz}, {Spitler},
  {Stansby}, {Streicher}, {{{S}}umak}, {Swinbank}, {Taranu}, {Tewary},
  {Tremblay}, {Val-Borro}, {Van Kooten}, {Vasovi{'c}}, {Verma}, {de Miranda
  Cardoso}, {Williams}, {Wilson}, {Winkel}, {Wood-Vasey}, {Xue}, {Yoachim},
  {Zhang}, {Zonca}, \& {Astropy Project Contributors}}]{astropy:2022}
{Astropy Collaboration}, {Price-Whelan}, A.~M., {Lim}, P.~L., {et~al.} 2022,
  apj, 935, 167, \dodoi{10.3847/1538-4357/ac7c74}

\bibitem[{{Ba{\~n}ados} {et~al.}(2018){Ba{\~n}ados}, {Venemans},
  {Mazzucchelli}, {Farina}, {Walter}, {Wang}, {Decarli}, {Stern}, {Fan},
  {Davies}, {Hennawi}, {Simcoe}, {Turner}, {Rix}, {Yang}, {Kelson}, {Rudie}, \&
  {Winters}}]{Banados2018Natur}
{Ba{\~n}ados}, E., {Venemans}, B.~P., {Mazzucchelli}, C., {et~al.} 2018, \nat,
  553, 473, \dodoi{10.1038/nature25180}

\bibitem[{{Barnett} {et~al.}(2017){Barnett}, {Warren}, {Becker}, {Mortlock},
  {Hewett}, {McMahon}, {Simpson}, \& {Venemans}}]{Barnett2017AA}
{Barnett}, R., {Warren}, S.~J., {Becker}, G.~D., {et~al.} 2017, \aap, 601, A16,
  \dodoi{10.1051/0004-6361/201630258}

\bibitem[{{Becker} {et~al.}(2015){Becker}, {Bolton}, {Madau}, {Pettini},
  {Ryan-Weber}, \& {Venemans}}]{Becker2015MNRAS}
{Becker}, G.~D., {Bolton}, J.~S., {Madau}, P., {et~al.} 2015, \mnras, 447,
  3402, \dodoi{10.1093/mnras/stu2646}

\bibitem[{{Becker} {et~al.}(2021){Becker}, {D'Aloisio}, {Christenson}, {Zhu},
  {Worseck}, \& {Bolton}}]{Becker2021MNRAS}
{Becker}, G.~D., {D'Aloisio}, A., {Christenson}, H.~M., {et~al.} 2021, \mnras,
  508, 1853, \dodoi{10.1093/mnras/stab2696}

\bibitem[{{Bosman} {et~al.}(2018){Bosman}, {Fan}, {Jiang}, {Reed}, {Matsuoka},
  {Becker}, \& {Haehnelt}}]{Bosman2018MNRAS}
{Bosman}, S. E.~I., {Fan}, X., {Jiang}, L., {et~al.} 2018, \mnras, 479, 1055,
  \dodoi{10.1093/mnras/sty1344}

\bibitem[{{Bosman} {et~al.}(2021){Bosman}, {{\v{D}}urov{\v{c}}{\'\i}kov{\'a}},
  {Davies}, \& {Eilers}}]{Bosman2021MNRAS}
{Bosman}, S. E.~I., {{\v{D}}urov{\v{c}}{\'\i}kov{\'a}}, D., {Davies}, F.~B., \&
  {Eilers}, A.-C. 2021, \mnras, 503, 2077, \dodoi{10.1093/mnras/stab572}

\bibitem[{{Calverley} {et~al.}(2011){Calverley}, {Becker}, {Haehnelt}, \&
  {Bolton}}]{Calverley2011MNRAS}
{Calverley}, A.~P., {Becker}, G.~D., {Haehnelt}, M.~G., \& {Bolton}, J.~S.
  2011, \mnras, 412, 2543, \dodoi{10.1111/j.1365-2966.2010.18072.x}

\bibitem[{{Carilli} {et~al.}(2010){Carilli}, {Wang}, {Fan}, {Walter}, {Kurk},
  {Riechers}, {Wagg}, {Hennawi}, {Jiang}, {Menten}, {Bertoldi}, {Strauss}, \&
  {Cox}}]{Carilli2010ApJ}
{Carilli}, C.~L., {Wang}, R., {Fan}, X., {et~al.} 2010, \apj, 714, 834,
  \dodoi{10.1088/0004-637X/714/1/834}

\bibitem[{{Davies} {et~al.}(2018){Davies}, {Hennawi}, {Ba{\~n}ados},
  {Luki{\'c}}, {Decarli}, {Fan}, {Farina}, {Mazzucchelli}, {Rix}, {Venemans},
  {Walter}, {Wang}, \& {Yang}}]{Davies2018ApJ}
{Davies}, F.~B., {Hennawi}, J.~F., {Ba{\~n}ados}, E., {et~al.} 2018, \apj, 864,
  142, \dodoi{10.3847/1538-4357/aad6dc}

\bibitem[{{Eilers} {et~al.}(2018){Eilers}, {Davies}, \&
  {Hennawi}}]{Eilers2018ApJ}
{Eilers}, A.-C., {Davies}, F.~B., \& {Hennawi}, J.~F. 2018, \apj, 864, 53,
  \dodoi{10.3847/1538-4357/aad4fd}

\bibitem[{{Eilers} {et~al.}(2017){Eilers}, {Davies}, {Hennawi}, {Prochaska},
  {Luki{\'c}}, \& {Mazzucchelli}}]{Eilers2017ApJ}
{Eilers}, A.-C., {Davies}, F.~B., {Hennawi}, J.~F., {et~al.} 2017, \apj, 840,
  24, \dodoi{10.3847/1538-4357/aa6c60}

\bibitem[{{Eilers} {et~al.}(2019){Eilers}, {Hennawi}, {Davies}, \&
  {O{\~n}orbe}}]{Eilers2019ApJ}
{Eilers}, A.-C., {Hennawi}, J.~F., {Davies}, F.~B., \& {O{\~n}orbe}, J. 2019,
  \apj, 881, 23, \dodoi{10.3847/1538-4357/ab2b3f}

\bibitem[{{Fan} {et~al.}(2006){Fan}, {Strauss}, {Becker}, {White}, {Gunn},
  {Knapp}, {Richards}, {Schneider}, {Brinkmann}, \& {Fukugita}}]{Fan2006AJ}
{Fan}, X., {Strauss}, M.~A., {Becker}, R.~H., {et~al.} 2006, \aj, 132, 117,
  \dodoi{10.1086/504836}

\bibitem[{{Finkelstein} {et~al.}(2019){Finkelstein}, {D'Aloisio},
  {Paardekooper}, {Ryan}, {Behroozi}, {Finlator}, {Livermore}, {Upton
  Sanderbeck}, {Dalla Vecchia}, \& {Khochfar}}]{Finkelstein2019ApJ}
{Finkelstein}, S.~L., {D'Aloisio}, A., {Paardekooper}, J.-P., {et~al.} 2019,
  \apj, 879, 36, \dodoi{10.3847/1538-4357/ab1ea8}

\bibitem[{{Greig} {et~al.}(2019){Greig}, {Mesinger}, \&
  {Ba{\~n}ados}}]{Greig2019MNRAS}
{Greig}, B., {Mesinger}, A., \& {Ba{\~n}ados}, E. 2019, \mnras, 484, 5094,
  \dodoi{10.1093/mnras/stz230}

\bibitem[{{Greig} {et~al.}(2022){Greig}, {Mesinger}, {Davies}, {Wang}, {Yang},
  \& {Hennawi}}]{Greig2022MNRAS}
{Greig}, B., {Mesinger}, A., {Davies}, F.~B., {et~al.} 2022, \mnras, 512, 5390,
  \dodoi{10.1093/mnras/stac825}

\bibitem[{{Greig} {et~al.}(2017){Greig}, {Mesinger}, {Haiman}, \&
  {Simcoe}}]{Greig2017MNRAS}
{Greig}, B., {Mesinger}, A., {Haiman}, Z., \& {Simcoe}, R.~A. 2017, \mnras,
  466, 4239, \dodoi{10.1093/mnras/stw3351}

\bibitem[{Harris {et~al.}(2020)Harris, Millman, van~der Walt, Gommers,
  Virtanen, Cournapeau, Wieser, Taylor, Berg, Smith, Kern, Picus, Hoyer, van
  Kerkwijk, Brett, Haldane, del R{\'{i}}o, Wiebe, Peterson,
  G{\'{e}}rard-Marchant, Sheppard, Reddy, Weckesser, Abbasi, Gohlke, \&
  Oliphant}]{Numpy2020}
Harris, C.~R., Millman, K.~J., van~der Walt, S.~J., {et~al.} 2020, Nature, 585,
  357, \dodoi{10.1038/s41586-020-2649-2}

\bibitem[{{Hoag} {et~al.}(2019){Hoag}, {Brada{\v{c}}}, {Huang}, {Mason},
  {Treu}, {Schmidt}, {Trenti}, {Strait}, {Lemaux}, {Finney}, \&
  {Paddock}}]{Hoag2019ApJ}
{Hoag}, A., {Brada{\v{c}}}, M., {Huang}, K., {et~al.} 2019, \apj, 878, 12,
  \dodoi{10.3847/1538-4357/ab1de7}

\bibitem[{Hunter(2007)}]{Matplotlib2007}
Hunter, J.~D. 2007, Computing in Science \& Engineering, 9, 90,
  \dodoi{10.1109/MCSE.2007.55}

\bibitem[{{Itoh} {et~al.}(2018){Itoh}, {Ouchi}, {Zhang}, {Inoue}, {Mawatari},
  {Shibuya}, {Harikane}, {Ono}, {Kusakabe}, {Shimasaku}, {Fujimoto}, {Iwata},
  {Kajisawa}, {Kashikawa}, {Kawanomoto}, {Komiyama}, {Lee}, {Nagao}, \&
  {Taniguchi}}]{Itoh2018ApJ}
{Itoh}, R., {Ouchi}, M., {Zhang}, H., {et~al.} 2018, \apj, 867, 46,
  \dodoi{10.3847/1538-4357/aadfe4}

\bibitem[{{Jung} {et~al.}(2020){Jung}, {Finkelstein}, {Dickinson}, {Hutchison},
  {Larson}, {Papovich}, {Pentericci}, {Straughn}, {Guo}, {Malhotra}, {Rhoads},
  {Song}, {Tilvi}, \& {Wold}}]{Jung2020ApJ}
{Jung}, I., {Finkelstein}, S.~L., {Dickinson}, M., {et~al.} 2020, \apj, 904,
  144, \dodoi{10.3847/1538-4357/abbd44}

\bibitem[{{Konno} {et~al.}(2014){Konno}, {Ouchi}, {Ono}, {Shimasaku},
  {Shibuya}, {Furusawa}, {Nakajima}, {Naito}, {Momose}, {Yuma}, \&
  {Iye}}]{Konno2014ApJ}
{Konno}, A., {Ouchi}, M., {Ono}, Y., {et~al.} 2014, \apj, 797, 16,
  \dodoi{10.1088/0004-637X/797/1/16}

\bibitem[{{Konno} {et~al.}(2018){Konno}, {Ouchi}, {Shibuya}, {Ono},
  {Shimasaku}, {Taniguchi}, {Nagao}, {Kobayashi}, {Kajisawa}, {Kashikawa},
  {Inoue}, {Oguri}, {Furusawa}, {Goto}, {Harikane}, {Higuchi}, {Komiyama},
  {Kusakabe}, {Miyazaki}, {Nakajima}, \& {Wang}}]{Konno2018PASJ}
{Konno}, A., {Ouchi}, M., {Shibuya}, T., {et~al.} 2018, \pasj, 70, S16,
  \dodoi{10.1093/pasj/psx131}

\bibitem[{{Mason} {et~al.}(2018){Mason}, {Treu}, {Dijkstra}, {Mesinger},
  {Trenti}, {Pentericci}, {de Barros}, \& {Vanzella}}]{Mason2018ApJ}
{Mason}, C.~A., {Treu}, T., {Dijkstra}, M., {et~al.} 2018, \apj, 856, 2,
  \dodoi{10.3847/1538-4357/aab0a7}

\bibitem[{{Mason} {et~al.}(2019){Mason}, {Fontana}, {Treu}, {Schmidt}, {Hoag},
  {Abramson}, {Amorin}, {Brada{\v{c}}}, {Guaita}, {Jones}, {Henry}, {Malkan},
  {Pentericci}, {Trenti}, \& {Vanzella}}]{Mason2019MNRAS}
{Mason}, C.~A., {Fontana}, A., {Treu}, T., {et~al.} 2019, \mnras, 485, 3947,
  \dodoi{10.1093/mnras/stz632}

\bibitem[{{McGreer} {et~al.}(2015){McGreer}, {Mesinger}, \&
  {D'Odorico}}]{McGreer2015MNRAS}
{McGreer}, I.~D., {Mesinger}, A., \& {D'Odorico}, V. 2015, \mnras, 447, 499,
  \dodoi{10.1093/mnras/stu2449}

\bibitem[{{McGreer} {et~al.}(2011){McGreer}, {Mesinger}, \&
  {Fan}}]{McGreer2011MNRAS}
{McGreer}, I.~D., {Mesinger}, A., \& {Fan}, X. 2011, \mnras, 415, 3237,
  \dodoi{10.1111/j.1365-2966.2011.18935.x}

\bibitem[{{Mesinger}(2010)}]{Mesinger2010MNRAS}
{Mesinger}, A. 2010, \mnras, 407, 1328,
  \dodoi{10.1111/j.1365-2966.2010.16995.x}

\bibitem[{{Millea} \& {Bouchet}(2018)}]{Millea2018AA}
{Millea}, M., \& {Bouchet}, F. 2018, \aap, 617, A96,
  \dodoi{10.1051/0004-6361/201833288}

\bibitem[{{Miralda-Escud{\'e}}(1998)}]{ME1998ApJ}
{Miralda-Escud{\'e}}, J. 1998, \apj, 501, 15, \dodoi{10.1086/305799}

\bibitem[{{Morales} {et~al.}(2021){Morales}, {Mason}, {Bruton}, {Gronke},
  {Haardt}, \& {Scarlata}}]{Morales2021ApJ}
{Morales}, A.~M., {Mason}, C.~A., {Bruton}, S., {et~al.} 2021, \apj, 919, 120,
  \dodoi{10.3847/1538-4357/ac1104}

\bibitem[{{Mortlock} {et~al.}(2011){Mortlock}, {Warren}, {Venemans}, {Patel},
  {Hewett}, {McMahon}, {Simpson}, {Theuns}, {Gonz{\'a}les-Solares}, {Adamson},
  {Dye}, {Hambly}, {Hirst}, {Irwin}, {Kuiper}, {Lawrence}, \&
  {R{\"o}ttgering}}]{Mortlock2011Natur}
{Mortlock}, D.~J., {Warren}, S.~J., {Venemans}, B.~P., {et~al.} 2011, \nat,
  474, 616, \dodoi{10.1038/nature10159}

\bibitem[{{Naidu} {et~al.}(2020){Naidu}, {Tacchella}, {Mason}, {Bose}, {Oesch},
  \& {Conroy}}]{Naidu2020ApJ}
{Naidu}, R.~P., {Tacchella}, S., {Mason}, C.~A., {et~al.} 2020, \apj, 892, 109,
  \dodoi{10.3847/1538-4357/ab7cc9}

\bibitem[{{Nasir} \& {D'Aloisio}(2020)}]{ND2020MNRAS}
{Nasir}, F., \& {D'Aloisio}, A. 2020, \mnras, 494, 3080,
  \dodoi{10.1093/mnras/staa894}

\bibitem[{{Osterbrock} {et~al.}(1996){Osterbrock}, {Fulbright}, {Martel},
  {Keane}, {Trager}, \& {Basri}}]{Osterbrock1996PASP}
{Osterbrock}, D.~E., {Fulbright}, J.~P., {Martel}, A.~R., {et~al.} 1996, \pasp,
  108, 277, \dodoi{10.1086/133722}

\bibitem[{{Ouchi} {et~al.}(2018){Ouchi}, {Harikane}, {Shibuya}, {Shimasaku},
  {Taniguchi}, {Konno}, {Kobayashi}, {Kajisawa}, {Nagao}, {Ono}, {Inoue},
  {Umemura}, {Mori}, {Hasegawa}, {Higuchi}, {Komiyama}, {Matsuda}, {Nakajima},
  {Saito}, \& {Wang}}]{Ouchi2018PASJ}
{Ouchi}, M., {Harikane}, Y., {Shibuya}, T., {et~al.} 2018, \pasj, 70, S13,
  \dodoi{10.1093/pasj/psx074}

\bibitem[{{Pentericci} {et~al.}(2011){Pentericci}, {Fontana}, {Vanzella},
  {Castellano}, {Grazian}, {Dijkstra}, {Boutsia}, {Cristiani}, {Dickinson},
  {Giallongo}, {Giavalisco}, {Maiolino}, {Moorwood}, {Paris}, \&
  {Santini}}]{Pentericci2011ApJ}
{Pentericci}, L., {Fontana}, A., {Vanzella}, E., {et~al.} 2011, \apj, 743, 132,
  \dodoi{10.1088/0004-637X/743/2/132}

\bibitem[{{Planck Collaboration} {et~al.}(2020){Planck Collaboration},
  {Aghanim}, {Akrami}, {Ashdown}, {Aumont}, {Baccigalupi}, {Ballardini},
  {Banday}, {Barreiro}, {Bartolo}, {Basak}, {Battye}, {Benabed}, {Bernard},
  {Bersanelli}, {Bielewicz}, {Bock}, {Bond}, {Borrill}, {Bouchet}, {Boulanger},
  {Bucher}, {Burigana}, {Butler}, {Calabrese}, {Cardoso}, {Carron},
  {Challinor}, {Chiang}, {Chluba}, {Colombo}, {Combet}, {Contreras}, {Crill},
  {Cuttaia}, {de Bernardis}, {de Zotti}, {Delabrouille}, {Delouis}, {Di
  Valentino}, {Diego}, {Dor{\'e}}, {Douspis}, {Ducout}, {Dupac}, {Dusini},
  {Efstathiou}, {Elsner}, {En{\ss}lin}, {Eriksen}, {Fantaye}, {Farhang},
  {Fergusson}, {Fernandez-Cobos}, {Finelli}, {Forastieri}, {Frailis},
  {Fraisse}, {Franceschi}, {Frolov}, {Galeotta}, {Galli}, {Ganga},
  {G{\'e}nova-Santos}, {Gerbino}, {Ghosh}, {Gonz{\'a}lez-Nuevo}, {G{\'o}rski},
  {Gratton}, {Gruppuso}, {Gudmundsson}, {Hamann}, {Handley}, {Hansen},
  {Herranz}, {Hildebrandt}, {Hivon}, {Huang}, {Jaffe}, {Jones}, {Karakci},
  {Keih{\"a}nen}, {Keskitalo}, {Kiiveri}, {Kim}, {Kisner}, {Knox},
  {Krachmalnicoff}, {Kunz}, {Kurki-Suonio}, {Lagache}, {Lamarre}, {Lasenby},
  {Lattanzi}, {Lawrence}, {Le Jeune}, {Lemos}, {Lesgourgues}, {Levrier},
  {Lewis}, {Liguori}, {Lilje}, {Lilley}, {Lindholm}, {L{\'o}pez-Caniego},
  {Lubin}, {Ma}, {Mac{\'\i}as-P{\'e}rez}, {Maggio}, {Maino}, {Mandolesi},
  {Mangilli}, {Marcos-Caballero}, {Maris}, {Martin}, {Martinelli},
  {Mart{\'\i}nez-Gonz{\'a}lez}, {Matarrese}, {Mauri}, {McEwen}, {Meinhold},
  {Melchiorri}, {Mennella}, {Migliaccio}, {Millea}, {Mitra},
  {Miville-Desch{\^e}nes}, {Molinari}, {Montier}, {Morgante}, {Moss}, {Natoli},
  {N{\o}rgaard-Nielsen}, {Pagano}, {Paoletti}, {Partridge}, {Patanchon},
  {Peiris}, {Perrotta}, {Pettorino}, {Piacentini}, {Polastri}, {Polenta},
  {Puget}, {Rachen}, {Reinecke}, {Remazeilles}, {Renzi}, {Rocha}, {Rosset},
  {Roudier}, {Rubi{\~n}o-Mart{\'\i}n}, {Ruiz-Granados}, {Salvati}, {Sandri},
  {Savelainen}, {Scott}, {Shellard}, {Sirignano}, {Sirri}, {Spencer},
  {Sunyaev}, {Suur-Uski}, {Tauber}, {Tavagnacco}, {Tenti}, {Toffolatti},
  {Tomasi}, {Trombetti}, {Valenziano}, {Valiviita}, {Van Tent}, {Vibert},
  {Vielva}, {Villa}, {Vittorio}, {Wandelt}, {Wehus}, {White}, {White},
  {Zacchei}, \& {Zonca}}]{Planck2020AA}
{Planck Collaboration}, {Aghanim}, N., {Akrami}, Y., {et~al.} 2020, \aap, 641,
  A6, \dodoi{10.1051/0004-6361/201833910}

\bibitem[{{Prochaska} {et~al.}(2020{\natexlab{a}}){Prochaska}, {Hennawi},
  {Westfall}, {Cooke}, {Wang}, {Hsyu}, {Davies}, {Farina}, \&
  {Pelliccia}}]{Prochaska2020JOSS}
{Prochaska}, J., {Hennawi}, J., {Westfall}, K., {et~al.} 2020{\natexlab{a}},
  The Journal of Open Source Software, 5, 2308, \dodoi{10.21105/joss.02308}

\bibitem[{{Prochaska} {et~al.}(2020{\natexlab{b}}){Prochaska}, {Hennawi},
  {Cooke}, {Westfall}, {Wang}, {EmAstro}, {Tiffanyhsyu}, {Wasserman},
  {Villaume}, {Marijana777}, {Schindler}, {Young}, {Simha}, {Wilde}, {Tejos},
  {Isbell}, {Fl{\"o}rs}, {Sandford}, {Vasovi{\'c}}, {Betts}, \&
  {Holden}}]{Prochaska2020zndo}
{Prochaska}, J.~X., {Hennawi}, J., {Cooke}, R., {et~al.} 2020{\natexlab{b}},
  {pypeit/PypeIt: Release 1.0.0}, v1.0.0,  Zenodo,
  \dodoi{10.5281/zenodo.3743493}

\bibitem[{{Robertson} {et~al.}(2015){Robertson}, {Ellis}, {Furlanetto}, \&
  {Dunlop}}]{Robertson2015ApJ}
{Robertson}, B.~E., {Ellis}, R.~S., {Furlanetto}, S.~R., \& {Dunlop}, J.~S.
  2015, \apjl, 802, L19, \dodoi{10.1088/2041-8205/802/2/L19}

\bibitem[{{Rousselot} {et~al.}(2000){Rousselot}, {Lidman}, {Cuby}, {Moreels},
  \& {Monnet}}]{Rousselot2000AA}
{Rousselot}, P., {Lidman}, C., {Cuby}, J.~G., {Moreels}, G., \& {Monnet}, G.
  2000, \aap, 354, 1134

\bibitem[{{Schenker} {et~al.}(2014){Schenker}, {Ellis}, {Konidaris}, \&
  {Stark}}]{Schenker2014ApJ}
{Schenker}, M.~A., {Ellis}, R.~S., {Konidaris}, N.~P., \& {Stark}, D.~P. 2014,
  \apj, 795, 20, \dodoi{10.1088/0004-637X/795/1/20}

\bibitem[{{Schroeder} {et~al.}(2013){Schroeder}, {Mesinger}, \&
  {Haiman}}]{SMH2013MNRAS}
{Schroeder}, J., {Mesinger}, A., \& {Haiman}, Z. 2013, \mnras, 428, 3058,
  \dodoi{10.1093/mnras/sts253}

\bibitem[{{Shull} {et~al.}(2012){Shull}, {Stevans}, \&
  {Danforth}}]{Shull2012ApJ}
{Shull}, J.~M., {Stevans}, M., \& {Danforth}, C.~W. 2012, \apj, 752, 162,
  \dodoi{10.1088/0004-637X/752/2/162}

\bibitem[{{Sobacchi} \& {Mesinger}(2015)}]{SM2015MNRAS}
{Sobacchi}, E., \& {Mesinger}, A. 2015, \mnras, 453, 1843,
  \dodoi{10.1093/mnras/stv1751}

\bibitem[{{Stark} {et~al.}(2010){Stark}, {Ellis}, {Chiu}, {Ouchi}, \&
  {Bunker}}]{Stark2010MNRAS}
{Stark}, D.~P., {Ellis}, R.~S., {Chiu}, K., {Ouchi}, M., \& {Bunker}, A. 2010,
  \mnras, 408, 1628, \dodoi{10.1111/j.1365-2966.2010.17227.x}

\bibitem[{{Venemans} {et~al.}(2015){Venemans}, {Ba{\~n}ados}, {Decarli},
  {Farina}, {Walter}, {Chambers}, {Fan}, {Rix}, {Schlafly}, {McMahon},
  {Simcoe}, {Stern}, {Burgett}, {Draper}, {Flewelling}, {Hodapp}, {Kaiser},
  {Magnier}, {Metcalfe}, {Morgan}, {Price}, {Tonry}, {Waters}, {AlSayyad},
  {Banerji}, {Chen}, {Gonz{\'a}lez-Solares}, {Greiner}, {Mazzucchelli},
  {McGreer}, {Miller}, {Reed}, \& {Sullivan}}]{Venemans2015ApJ}
{Venemans}, B.~P., {Ba{\~n}ados}, E., {Decarli}, R., {et~al.} 2015, \apjl, 801,
  L11, \dodoi{10.1088/2041-8205/801/1/L11}

\bibitem[{Virtanen {et~al.}(2020)Virtanen, Gommers, Oliphant, Haberland, Reddy,
  Cournapeau, Burovski, Peterson, Weckesser, Bright, {van der Walt}, Brett,
  Wilson, Millman, Mayorov, Nelson, Jones, Kern, Larson, Carey, Polat, Feng,
  Moore, {VanderPlas}, Laxalde, Perktold, Cimrman, Henriksen, Quintero, Harris,
  Archibald, Ribeiro, Pedregosa, {van Mulbregt}, \& {SciPy 1.0
  Contributors}}]{2020SciPy-NMeth}
Virtanen, P., Gommers, R., Oliphant, T.~E., {et~al.} 2020, Nature Methods, 17,
  261, \dodoi{10.1038/s41592-019-0686-2}

\bibitem[{{Wang} {et~al.}(2020){Wang}, {Davies}, {Yang}, {Hennawi}, {Fan},
  {Barth}, {Jiang}, {Wu}, {Mudd}, {Ba{\~n}ados}, {Bian}, {Decarli}, {Eilers},
  {Farina}, {Venemans}, {Walter}, \& {Yue}}]{Wang2020ApJ}
{Wang}, F., {Davies}, F.~B., {Yang}, J., {et~al.} 2020, \apj, 896, 23,
  \dodoi{10.3847/1538-4357/ab8c45}

\bibitem[{{Worseck} {et~al.}(2014){Worseck}, {Prochaska}, {O'Meara}, {Becker},
  {Ellison}, {Lopez}, {Meiksin}, {M{\'e}nard}, {Murphy}, \&
  {Fumagalli}}]{Worseck2014MNRAS}
{Worseck}, G., {Prochaska}, J.~X., {O'Meara}, J.~M., {et~al.} 2014, \mnras,
  445, 1745, \dodoi{10.1093/mnras/stu1827}

\bibitem[{{Yang} {et~al.}(2020{\natexlab{a}}){Yang}, {Wang}, {Fan}, {Hennawi},
  {Davies}, {Yue}, {Eilers}, {Farina}, {Wu}, {Bian}, {Pacucci}, \&
  {Lee}}]{Yang2020ApJb}
{Yang}, J., {Wang}, F., {Fan}, X., {et~al.} 2020{\natexlab{a}}, \apj, 904, 26,
  \dodoi{10.3847/1538-4357/abbc1b}

\bibitem[{{Yang} {et~al.}(2020{\natexlab{b}}){Yang}, {Wang}, {Fan}, {Hennawi},
  {Davies}, {Yue}, {Banados}, {Wu}, {Venemans}, {Barth}, {Bian}, {Boutsia},
  {Decarli}, {Farina}, {Green}, {Jiang}, {Li}, {Mazzucchelli}, \&
  {Walter}}]{Yang2020ApJa}
---. 2020{\natexlab{b}}, \apjl, 897, L14, \dodoi{10.3847/2041-8213/ab9c26}

\bibitem[{{Zhu} {et~al.}(2021){Zhu}, {Becker}, {Bosman}, {Keating},
  {Christenson}, {Ba{\~n}ados}, {Bian}, {Davies}, {D'Odorico}, {Eilers}, {Fan},
  {Haehnelt}, {Kulkarni}, {Pallottini}, {Qin}, {Wang}, \& {Yang}}]{Zhu2021ApJ}
{Zhu}, Y., {Becker}, G.~D., {Bosman}, S. E.~I., {et~al.} 2021, \apj, 923, 223,
  \dodoi{10.3847/1538-4357/ac26c2}

\bibitem[{{Zhu} {et~al.}(2022){Zhu}, {Becker}, {Bosman}, {Keating},
  {D'Odorico}, {Davies}, {Christenson}, {Ba{\~n}ados}, {Bian}, {Bischetti},
  {Chen}, {Davies}, {Eilers}, {Fan}, {Gaikwad}, {Greig}, {Haehnelt},
  {Kulkarni}, {Lai}, {Pallottini}, {Qin}, {Ryan-Weber}, {Walter}, {Wang}, \&
  {Yang}}]{Zhu2022ApJ}
---. 2022, \apj, 932, 76, \dodoi{10.3847/1538-4357/ac6e60}

\end{thebibliography}
\bibliographystyle{aasjournal}



\end{document}